# On inverse aggregate network assignment games

**Joseph Y. J. Chow**
C2SMART Center, Department of Civil, Urban, and Environmental Engineering,
New York University Tandon School of Engineering, Brooklyn, NY, USA

## Abstract
Multi-operator mobility markets exist in many forms: airlines, freight, Mobility-as-a-Service systems, agency coordination in humanitarian logistics, shared autonomous vehicle fleet operations, and more. To design platforms coordinating their operations or infrastructure to support them, models are needed to understand the behavior of these operators and travelers in a shared network. Assignment games can prove useful, but operationalizing them requires adopting a more generalized aggregate assignment game framework that allows for identification and estimation. We extend such a framework to aggregate network assignment games, proposing a path-based formulation and a PURC-based formulation along with solution algorithms. An inverse optimization model is proposed for the link-based formulation. It is solved by a quadratic program that separates into a buyer block, which retains the node potentials and imposes every observed zero-flow cell, and a seller block. This estimator, introduced in the literature for uncapacitated route choice, is extended here to a capacitated two-sided market, with rationing prices supported on observed binding links and operator supply scales estimated by group. Computational tests with a toy network demonstrate the economic value of these models. A large-scale Long Island Rail Road network example (897 nodes, 671 census tracts, 123 LIRR stations, 8,320 directed links) is calibrated to 2026 fares, timetabled seat supply and National Transit Database operating costs. With 10-50% of links sampled, the misspecified inverse model fits oracle link flows with $R^2 \geq 96\%$. Held-out link error falls from 29% to 19%, correlations with the true rationing prices rise from 49% to 75%, and social optimal gross surplus is recovered within 6%. Deadweight loss is overestimated by 10-17%. Ablation tests show that ignoring the inactive cells more than doubles the held-out error and, at 30-50% sampling, roughly halves the correlation of the pricing wedge, while ignoring the rationing prices degrades every metric.

## 1. Introduction
For decades, there has been a need to understand how different mobility service providers interact in a network setting. Deregulation of airlines and trucking in the 1970s to 1980s (e.g. Airline Deregulation Act of 1978, Motor Carrier Act of 1980 in the U.S.) prompted the study of multi-operator mobility service networks in airlines and freight. These include noncooperative games in airlines, e.g. airline oligopolies (Reiss and Spiller, 1989; Brander and Zhang, 1990) or hub dominance and network competition (Hendricks et al., 1997); in freight and supply chains, e.g. (Harker and Friesz, 1986a,b; Friesz et al., 1986; Nagurney et al., 2002; Carr and Karmarkar, 2005); and cooperative games of alliance formation or cost transfers: in airlines (Littlechild and Owen, 1973; Brueckner, 2001; Park, 1997) and in freight (Krajewska et al., 2008; Özener and Ergun, 2008; Agarwal and Ergun, 2010). Urban transit services have seen less attention given the regional dominance of transit agencies, although in some markets (e.g. Hong Kong), there are similar studies (Zhou et al., 2005).

In recent years, advances in information and communications technologies have led to the emergence of private mobility service providers in urban markets, with microtransit/ride-pooling, ride-hail, carshare, micromobility, and other Mobility-on-Demand technologies (Djavadian and Chow, 2017). These advances garnered even greater attention from researchers interested in multi-operator interactions in these contexts, whether they are noncooperative (Yang et al., 2002; Zha et al., 2016; Zhou et al., 2022; Xi et al., 2024; Yao et al., 2026) or cooperative/coopetitive (Rasulkhani and Chow, 2019; Pantelidis et al., 2020; Liu and Chow, 2024; Yang and Chow, 2026). Similar multi-operator mobility service systems gaining traction include agency coordination for humanitarian logistics, robot swarm or autonomous vehicle fleet operations, and integrated electric mobility systems.

We consider cooperative network games which can answer the following types of questions. Under what incentives can multiple operators and travelers come together to make a mobility service market or platform economically stable, considering selfish behaviors and ability to form alliances? How can different network infrastructure (hubs, link capacities, service region coverage) affect the market equilibrium? What are the effects of operator decisions (e.g. entry/exit, new investments) that impact other operators and travelers?

An assignment game (Shapley and Shubik, 1971) is used to model the exchange of incentives between players, i.e. cost allocation, between the different players split as either buyers (travelers) or sellers (mobility operators) in a two-sided market. It is the transferable-utility counterpart of a stable matching problem, i.e. the marriage problem (Gale and Shapley, 1962; Becker, 1973, 1974), where matches are accompanied by transferable utilities which can reflect fares, subsidies, or tolls. The solution of the problem is a set of optimal matches and the cost transfers required to ensure those matches are stable as a socially optimal market equilibrium. The original studies consider only buyers and sellers, not the more complex network settings involving travel paths and link capacities that impose rationing prices on travelers to divert them to other routes.

In the network context, Kalai and Zemel (1982) introduce flow games, in which players own links and a coalition's value is the maximum flow it can route over the links it owns. Agarwal and Ergun (2008) examine multicommodity flow games for designing service network alliances, using inverse optimization (Ahuja and Orlin, 2001) to construct a vector of internal transfer prices between operators to achieve desired flows as optimal. These are one-sided games where transfers are only between operators. On the other hand, Liu and Chow (2024) and Yang and Chow (2026) extend assignment games to such *network assignment games* with cost transfers between travelers and operators and allow for subsidies between operators and the regulatory agency. This approach can be applied to design alliances with fare bundle design in airlines or freight, plan infrastructure for service systems, design platforms for Mobility-as-a-Service, among others.

Operationalizing network assignment games requires a method to fit the parameters to empirical data, i.e. the inverse problem (Tarantola, 2005). Researchers have proposed a modification to assignment games to make them identifiable. *Aggregate assignment games* model which buyers and sellers are matched using random utility models to fit observable attributes to characterize the payoffs using dual entropy formulations (Galichon and Salanié, 2022). The original assignment game is shown to be a limiting case. In a network context, however, observability can be a primary issue, where traveler path flows may not be easily observed.

Several contributions are made in this work. First, the aggregate assignment game is formally extended to an aggregate network assignment game (P1) using the dual entropy formulation framework from Galichon and Salanié (2022). Travelers are matched along paths in the network, while operators allocate capacities to links they own serving different paths. Second, to reduce the data requirements from path flows to sampled segment-link flows plus aggregate link counts, we modify the aggregate network assignment game into a link-based formulation and establish its relationship to P1 and its properties. Third, an inverse optimization model is proposed, i.e. the inverse aggregate network assignment game. It combines samples of traveler segment-link volumes and prior link operating costs with aggregate link counts and capacities. From these it infers the attributes of a Perturbed Utility Route Choice model for travelers and $L_1$-norm perturbed link costs from a prior for operators. Fourth, we show that this inverse problem can be solved as two convex quadratic programs. The buyer block retains the node potentials and imposes every observed inactive cell as a constraint, following Rosenblad et al. (2026), who introduced this treatment for uncapacitated route choice. We extend it to a capacitated two-sided market in three ways:

- Rationing prices are supported only on links observed at capacity and are bounded by the observed fares;
- A seller block recovers operating costs and supply scales shared within operator groups;

- Estimation is followed by fixed-price and clearing solves that recover the gross surplus, deadweight loss and pricing wedge.

Computational experiments on a large-scale synthetic Long Island Rail Road (LIRR) network, built from published fares, timetabled seat supply and transit cost data, validate the properties of the models and algorithms.

## 2. Background and related work

An overview of the existing literature of the inverse optimization problem starting from one-to-one assignment games to aggregate assignment games is provided.

### *2.1. Inverse one-to-one assignment games*

We adopt the notation of Shapley and Shubik (1971) for this section. A set of buyers and sellers with a known vector of payoffs under different matches are matched while identifying the corresponding cost transfers between them needed to stabilize those matches. This serves as a generalization of the basic assignment problem, which only outputs the matches.

Let
$M$ be the set of buyers
$N$ be the set of sellers
$a_{ij}$ = payoff from matching $(i, j)$
$x_{ij}$ be 1 if there is a match between $(i, j)$, 0 otherwise

$$\max_{x} z = \sum_{i \in M} \sum_{j \in N} a_{ij} x_{ij} \tag{1}$$

s.t.

$$\sum_{i \in M} x_{ij} \leq 1, \qquad \forall j \in N \tag{2}$$

$$\sum_{j \in N} x_{ij} \leq 1, \qquad \forall i \in M \tag{3}$$

$$x_{ij} \geq 0 \tag{4}$$

Since the constraint matrix is totally unimodular, it can be solved in relaxed continuous form (Eq. (4)) to obtain binary variables. The core of this assignment game determines the set of cost transfers that form a stable outcome. In the one-to-one assignment game, the core is defined by dual variables shown in Eq. (5) – (6).

$$\min_{u \geq 0, v \geq 0} w = \sum_{i \in M} u_i + \sum_{j \in N} v_j \tag{5}$$

s.t.

$$u_i + v_j \geq a_{ij}, \qquad \forall i \in M, j \in N \tag{6}$$

In inverse optimization, a solution of a forward model is observed, and the parameters are estimated from a prior such that the solution would be optimal. Inverse optimization has been applied to traffic assignment (Chow et al., 2014; Liu and Yin, 2026), vehicle routing (Chow and Recker, 2012; You et al., 2016; Chen et al., 2021; Zattoni Scroccaro et al., 2025), scheduling (Brucker and Shakhlevich, 2009), and facility location (Burkard et al., 2004), among others. For the inverse one-to-one assignment game, the matches are observed, $x_{ij}^0$. In a conventional *assignment problem*, a centralized decision-maker would also have access to a prior payoff vector $a_{ij}^0$ from which a set of $a_{ij}$ are determined such that the observed $x_{ij}^0$ are optimal. This is the inverse assignment problem (Ahuja and Orlin, 2001, for the discrete assignment, or the continuous assignment case as a Monge-Kantorovich optimal transport problem with Stuart and Wolfram, 2020).

However, in an *assignment game* that considers decentralized agents, the $a_{ij}$ are truly latent and not known even as a prior, making the inverse problem underdetermined. For example, a match might result not because the two prefer each other the most, but simply because the preferred partners are already matched elsewhere and the current match has a higher payoff than staying single.

### 2.2. Inverse aggregate assignment games

This led researchers to generalize the assignment game to a form that is identifiable, by relating the latent payoffs to observable attributes using discrete choice modeling mechanisms to describe aggregate matches. Dagsvik (2000) proposed the idea of aggregate matching markets with nontransferable utilities by incorporating a logit function as unobserved shocks to feed into a matching algorithm. Choo and Siow (2006) use that logit function explicitly to model aggregate choices of buyers and sellers to achieve an equilibrium matching that can be estimated and has a unique global optimum, which we call an *aggregate assignment game*. Galichon and Salanié (2022) showed that the aggregate assignment game is an entropy-regularized version of the Shapley-Shubik assignment game.

Let

The set of buyers and sellers be indexed by type, i.e. $M_i$ is the set of buyers of type $i$, and $N_j$ is the set of sellers of type $j$,

$\mu_{ij}$ denote the number of matches between $(i, j)$, where $\mu_{i0}$ is the number of unmatched buyers and $\mu_{0j}$ the number of unmatched sellers.

Buyers have a utility function in Eq. (7).

$$U_{ijg} = a_{ij} - \tau_{ij} + \epsilon_{ijg} \tag{7}$$

$$U_{i0g} = a_{i0} + \epsilon_{i0g} \tag{8}$$

where $a_{ij}$ is an alternative specific constant for utility gained from the match, $\tau_{ij}$ is the cost transfer from buyer to seller, and $\epsilon_{ijg}$ is an i.i.d. Gumbel variate for each buyer $g$ of type $i$ matching with a seller type $j$. In Eq. (8), $a_{i0}$ ($\epsilon_{i0g}$) is the utility (disturbance) from having no matches.

Sellers have a similar utility function in Eq. (9) with seller $h$ of type $j$ matching with type $i$ buyers.

$$V_{ijh} = b_{ij} + \tau_{ij} + \eta_{ijh} \tag{9}$$

$$V_{0jh} = b_{0j} + \eta_{0jh} \tag{10}$$

The logit-based random utility of each buyer's (seller's) choice relative to an option of no match is set as a log ratio in Eq. (11) (Eq. (12)):

$$\ln\left(\frac{\mu_{ij}}{\mu_{i0}}\right) = a_{ij} - a_{i0} - \tau_{ij} \tag{11}$$

$$\ln\left(\frac{\mu_{ij}}{\mu_{0j}}\right) = b_{ij} - b_{0j} + \tau_{ij} \tag{12}$$

By defining $\alpha_{ij}$ as the social surplus of this two-sided market, we get Eq. (13).

$$\alpha_{ij} = \ln\left(\frac{\mu_{ij}}{\mu_{i0}}\right) + \ln\left(\frac{\mu_{ij}}{\mu_{0j}}\right) = a_{ij} + b_{ij} - a_{i0} - b_{0j} \tag{13}$$

The authors show this yields Eq. (14).

$$\mu_{ij} = \sqrt{\mu_{i0}\mu_{0j}} \exp\left(\frac{\alpha_{ij}}{2}\right) \tag{14}$$

As a mathematical programming model, Galichon and Salanié (2022) show this to be Eq. (15) – (18).

$$\max_{\boldsymbol{\mu}} \left( \sum_{(i,j)} \mu_{ij}\alpha_{ij} + \sigma \cdot \mathcal{E}(\boldsymbol{\mu}) \right) \tag{15}$$

s.t.

$$\sum_j \mu_{ij} + \mu_{i0} = M_i, \qquad \forall i \tag{16}$$

$$\sum_i \mu_{ij} + \mu_{0j} = N_j, \qquad \forall j \tag{17}$$

$$\boldsymbol{\mu} \geq \mathbf{0} \tag{18}$$

where $\mathcal{E}(\boldsymbol{\mu})$ is a generalized entropy that can be represented as a Shannon entropy, and $\sigma > 0$ is a scale parameter for the noise. When $\sigma \to 0$, the logit scale parameter $\frac{1}{\sigma} \to \infty$, and the entropy regularization term disappears, reverting to the original formulation from Shapley and Shubik (1971).

The inverse model for Eqs. (15) – (18) determines payoff shares nonparametrically given $\mu^0$. The entropy term is strictly concave, creating a one-to-one mapping between counts $\boldsymbol{\mu}$ and surplus $\boldsymbol{\alpha}$. As a result, the inverse aggregate matching model is not an optimization, it is simply solving for the values of $\boldsymbol{\alpha}$ using the gradients. The corresponding estimate of the latent surplus $\alpha_{ij}$ is then captured with Eq. (19).

$$\hat{\alpha}_{ij} = 2\ln(\mu_{ij}) - \ln(\mu_{i0}) - \ln(\mu_{0j}) \tag{19}$$

Liu et al. (2025) propose an alternative model paradigm, considering a coalition-based set of decision-makers instead of modeling each of the buyers' and sellers' choices separately with random utility models.

## 3. Aggregate network assignment games

For the sake of clarity, we reset the notation from here on.

### *3.1. Path flow formulation*

Let

$Q$: set of buyers (traveler OD segments) with total demand $D_i, i \in Q$

$K_i$: set of candidate paths connecting OD segment $i$, where $K = \bigcup_{i \in Q} K_i$

$M$: set of sellers (operators)

$A_m$: set of links owned by operator $m$, where $A = A_0 \cup \{\bigcup_{m \in M} A_m\}$, where $A_0$ is the set of links not owned by any operator (including outside options directly connecting each OD pair), and $a_i^0 \in A_0$ is the outside link directly connecting the OD pair for segment $i \in Q$, and $k_i^0 \in K_i$ is an outside path that consists solely of $a_i^0$

$A_k$: set of links belonging to path $k \in K$

$A^{own} = \bigcup_{m \in M} A_m$: set of links owned by operators

$\delta_{ak}$: link-path incidence, set to 1 if link $a \in A$ is on path $k \in K$, 0 otherwise

$C_a^{max}$: maximum flow capacity limit for link $a \in A_m, m \in M$

$F_m$: capacity available to operator $m \in M$, where $F_m < \sum_{a \in A_m} C_a^{max}$

$T_k \equiv \sum_{a \in A_k \cap A^{own}} \tau_a$: path price of path $k \in K$, the sum of cost transfers on its owned links; $T_{k_i^0} = 0$.

$c_a$: unit cost of providing capacity to link $a \in A_m$ for operator $m \in M$

$\Phi_{ik}$: gross deterministic joint surplus derived when traveler $i \in Q$ uses path $k \in K_i$ *relative to not traveling on the network*, where $\Phi_{i0} \equiv \Phi_{ik_i^0}$ is the gross surplus of the outside option, normalized to $\Phi_{i0} = 0$ for all $i \in Q$.
$\alpha_B > 0$ ($\alpha_{S,m} > 0$): logit scale parameter governing traveler path choice (operator $m$'s capacity allocation choice considering noise from information unavailability) dispersion; $\alpha \to \infty$ recovers the deterministic choice and corresponds to Shapley-Shubik's deterministic assignment game structure.
$x_{ik} \geq 0$: path matching flow for traveler OD segment $i \in Q$ on path $k \in K$, where $x_{ik_i^0}$ is flow on an outside path
$y_a \geq 0$: continuous capacity allocated by operator $m \in M$ to link $a \in A_m$, where $y_{m0}$ is reserve capacity with cost $c_{idle,m}$

In this study, we follow the convention of Galichon and Salanié (2022) to derive a two-sided entropy, single level aggregate network assignment game under a multi-operator path network subject to operator capacity budget constraints. A market is served by a set of operators $M$ deciding how much of their capacities $F_m$ are allocated to different links that they own. Travelers in each segment $i \in Q$ decide which path to take to make their trip. The matching is the combination of path flows allowed by the combined link capacities allocated from one or more operators. In this aggregate network assignment game formulation, travelers choose in aggregate among a set of paths with different probabilities based on the utilities they attain with each path. Operators choose the shares of their capacities to assign to each link that they own based on the utilities they gain from each link. Optimal cost transfers are determined through the tolls that operators charge to travelers on the links.

Compared to Liu et al. (2025), this has bi-regularized entropy and we solve for a first-best optimal pricing endogenous to the social optimum. By comparison, Liu et al. (2025) solve a second best pricing strategy as a bilevel problem, where the assignment occurs in the lower level with cost transfers fixed from the upper level.

P1:

$$\max_{x,y} Z = \sum_{i\in Q}\sum_{k\in K_i}\Phi_{ik}x_{ik} - \sum_{m\in M}\sum_{a\in A_m} c_a y_a - \sum_{m\in M} c_{idle,m}y_{m0} - \frac{1}{\alpha_B}\sum_{i\in Q}\sum_{k\in K_i} x_{ik}(\ln x_{ik} - 1) - \sum_{m\in M}\frac{1}{\alpha_{S,m}}\left(\sum_{a\in A_m} y_a(\ln y_a - 1) + y_{m0}(\ln y_{m0} - 1)\right) \tag{20}$$

s.t.

Buyer demand conservation:

$$\sum_{k\in K_i\setminus\{k_i^0\}} x_{ik} + x_{ik_i^0} = D_i, \qquad \forall i \in Q \qquad (u_i) \tag{21}$$

Link market clearing:

$$\sum_{i\in Q}\sum_{k\in K_i}\delta_{ak}x_{ik} \leq y_a, \qquad \forall a \in A_m, m \in M \qquad (\tau_a) \tag{22}$$

Link flow capacity constraints:

$$y_a \leq C_a^{max}, \qquad \forall\, a \in A_m, m \in M \qquad (\pi_a) \tag{23}$$

Budget constraint:

$$\sum_{a\in A_m} y_a + y_{m0} = F_m, \qquad \forall m \in M \qquad (v_m) \tag{24}$$

Non-negativity:

$$x_{ik} \geq 0, \qquad \forall i \in Q, k \in K_i$$
$$y_a \geq 0, \qquad \forall a \in A_m, m \in M \tag{25}$$

$$y_{m0} \geq 0, \qquad \forall m \in M$$

The outside links $a_i^0$ carry no market clearing constraint (Eq. (22)), no flow capacity constraint (Eq. (23)), and no cost transfer, so $T_{k_i^0} = 0$. The objective $Z$ itself is not the social surplus due to the dropped constants; instead, the social surplus is shown in Eq. (20a).

$$\tilde{Z}^* = Z^* + \frac{1}{\alpha_B}\left(\sum_{i\in Q} D_i \ln D_i - \sum_{i\in Q} D_i\right) + \sum_{m\in M} \frac{1}{\alpha_{S,m}}(F_m \ln F_m - F_m) \tag{20a}$$

We can take the Lagrangian $\mathcal{L}$ in Eq. (26) and derive the following first order necessary conditions.

$$\mathcal{L} = Z(\boldsymbol{x},\boldsymbol{y}) - \sum_{i\in Q} u_i\left(\sum_{k\in K_i} x_{ik} - D_i\right) - \sum_{m\in M}\sum_{a\in A_m} \tau_a\left(\sum_{i\in Q}\sum_{k\in K_i} \delta_{ak}x_{ik} - y_a\right)$$
$$- \sum_{m\in M}\sum_{a\in A_m} \pi_a(y_a - C_a^{max}) - \sum_{m\in M} v_m\left(\sum_{a\in A_m} y_a + y_{m0} - F_m\right) \tag{26}$$

Determining buyer path choice by taking $\frac{\partial \mathcal{L}}{\partial x_{ik}} = 0$:

$$\Phi_{ik} - \frac{1}{\alpha_B}\ln x_{ik}^* - u_i^* - T_k^* = 0 \tag{27}$$

For $\Phi_{i0} = 0$, the cost transfer sum vanishes, leading to $-\frac{1}{\alpha_B}\ln x_{i0}^* - u_i^* = 0$, which means $x_{i0}^* = \exp(-\alpha_B u_i^*)$, $u_i^* = -\frac{1}{\alpha_B}\ln x_{i0}^*$.

Solving for the dual variables we get the following relationships:

$$x_{ik}^* = D_i \frac{\exp\big(\alpha_B(\Phi_{ik} - T_k^*)\big)}{1 + \sum_{k'\in K_i\setminus\{k_i^0\}} \exp\Big(\alpha_B\big(\Phi_{ik'} - T_{k'}^*\big)\Big)} \tag{28}$$

$$u_i^* = \frac{1}{\alpha_B}\ln\left(1 + \sum_{k\in K_i\setminus\{k_i^0\}} \exp\big(\alpha_B(\Phi_{ik} - T_k^*)\big)\right) - \frac{1}{\alpha_B}\ln D_i \tag{29}$$

$u_i^*$ is the shadow price of segment $i$'s demand conservation constraint. It mixes the inclusive value of $i$'s path set with segment size through the $-\frac{1}{\alpha_B}\ln D_i$ term, so we define the recentered dual $\tilde{u}_i \equiv u_i + \frac{1}{\alpha_B}\ln D_i \geq 0$, which is the inclusive value alone and is therefore comparable across segments of different size, and is the expected equilibrium utility.

Determining seller capacity provision by taking $\frac{\partial \mathcal{L}}{\partial y_a} = 0$:

$$-c_a - \frac{1}{\alpha_{S,m}}\ln y_a^* + \tau_a^* - \pi_a^* - v_m^* = 0 \tag{30}$$

Solving for $y_a^*$, we get an exponential link supply function:

$$y_a^* = \exp\Big(\alpha_{S,m}(\tau_a^* - c_a - \pi_a^* - v_m^*)\Big) \tag{31}$$

where the optimal reserve allocation is $y_{m0}^* = \exp\left(-\alpha_{S,m}\left(c_{idle,m} + v_m^*\right)\right)$. The $v_m^*$ is the shadow price of operator $m$'s capacity budget, where $v_m^* = \frac{\partial Z^*}{\partial F_m}$. Substituting Eq. (31) into the budget equality constraint Eq. (24) and taking natural logs, we get the following expression in Eq. (32) for $v_m^*$.

$$v_m^* = \frac{1}{\alpha_{S,m}} \ln\left( \exp\left(-\alpha_{S,m} c_{idle,m}\right) + \sum_{a \in A_m} \exp\left(\alpha_{S,m}(\tau_a^* - c_a - \pi_a^*)\right) \right) - \frac{1}{\alpha_{S,m}} \ln F_m \tag{32}$$

As with the buyer side, $v_m^*$ carries a scale term. We define $\tilde{v}_m \equiv v_m + \frac{1}{\alpha_{S,m}} \ln F_m \geq -c_{idle,m}$, which by Eq. (32) is the operator's inclusive value $\frac{1}{\alpha_{S,m}} \ln\left(\exp\left(-\alpha_{S,m} c_{idle,m}\right) + \sum_{a \in A_m} \exp\left(\alpha_{S,m}(\tau_a^* - c_a - \pi_a^*)\right)\right)$ and is free of $F_m$. Differentiating Eq. (20a), $\frac{\partial \tilde{Z}^*}{\partial F_m} = v_m^* + \frac{1}{\alpha_{S,m}} \ln F_m = \tilde{v}_m$, so $\tilde{v}_m$ is the marginal social surplus of one more unit of endowment. Two operators $m$ and $m'$ identical at the margin but differing in endowment share the same $\tilde{v}_m$ while their $v_m$ differ by $\frac{1}{\alpha_{S,m'}} \ln(F_{m'}) - \frac{1}{\alpha_{S,m}} \ln(F_m)$. Welfare comparisons across operators must be made in $\tilde{v}_m$.

Plugging this back into Eq. (31) gets the logit function for the operator $m \in M$ in Eq. (33) for allocating capacity to a link $a \in A_m$.

$$y_a^* = F_m \frac{\exp\left(\alpha_{S,m}(\tau_a^* - c_a - \pi_a^*)\right)}{\exp\left(-\alpha_{S,m} c_{idle,m}\right) + \sum_{a' \in A_m} \exp\left(\alpha_{S,m}(\tau_{a'}^* - c_{a'} - \pi_{a'}^*)\right)} \tag{33}$$

**Assumption 1 (regularity)**. *For every $m \in M$: $F_m > 0, C_a^{max} > 0$ for all $a \in A_m$, and $F_m < \sum_{a \in A_m} C_a^{max}$, so that the budget cannot exceed total physical capacity. For every $i \in Q$: $D_i > 0$, and $K_i$ contains the outside path $k_i^0$.*

**Theorem 1. Duality and correspondence to bi-regularized assignment games.** *Let the proposed problem in Eq. (20) – (25) satisfy Assumption 1, defined over the convex feasible set of path flows $\boldsymbol{x}$ and continuous link capacities $\boldsymbol{y}$ with two-sided Shannon entropy regularization. Then the network equilibrium satisfies the following:*

1. ***Box-constrained dual representation**. The Lagrangian dual $W(\boldsymbol{u}, \boldsymbol{\tau}, \boldsymbol{\pi}, \boldsymbol{v})$ of Eqs. (20) – (25) is jointly convex on $\{\boldsymbol{u} \in \mathbb{R}^{|Q|}, \boldsymbol{\tau} \geq 0, \boldsymbol{\pi} \geq 0, \boldsymbol{v} \in \mathbb{R}^{|M|}\}$, and strong duality holds with zero duality gap. Because $u_i$ and $v_m$ admit the closed forms in Eq. (29) and Eq. (32), they may be concentrated out analytically, yielding the box-constrained convex program $\mathcal{W}$ in Eq. (34) over $(\boldsymbol{\tau}, \boldsymbol{\pi})$ alone, in which the demand conservation constraints in Eq. (21) and the capacity budget in Eq. (24) are satisfied identically by construction.*

$$\begin{aligned} \min_{\tau \geq 0, \pi \geq 0} \mathcal{W} = & \frac{1}{\alpha_B} \sum_{i \in Q} D_i \left( ln\left( 1 + \sum_{k \in K_i \setminus \{k_i^0\}} \exp\left(\alpha_B(\Phi_{ik} - T_k)\right) \right) - \ln D_i + 1 \right) \\ & + \sum_{m \in M} \frac{1}{\alpha_{S,m}} F_m \left( \ln\left( \exp\left(-\alpha_{S,m} c_{idle,m}\right) + \sum_{a \in A_m} \exp\left(\alpha_{S,m}(\tau_a - c_a - \pi_a)\right) \right) - \ln F_m \right. \\ & \left. + 1 \right) + \sum_{m \in M} \sum_{a \in A_m} \pi_a C_a^{max} \end{aligned} \tag{34}$$

2. ***Primal-dual isomorphism and dual determinacy**. $Z$ is strictly concave over a nonempty compact feasible set, so the primal optimum $(\boldsymbol{x}^*, \boldsymbol{y}^*)$ exists and is unique. Two dual optima differ by a perturbation $(\boldsymbol{d\tau}, \boldsymbol{d\pi})$ satisfying simultaneously*

$$\sum_{a \in A_k \cap A^{own}} d\tau_a = 0 \ \forall k, \qquad \boldsymbol{d\pi} = \boldsymbol{d\tau}, \qquad \sum_{m \in M} \sum_{a \in A_m} d\tau_a C_a^{max} = 0,$$

*while keeping $\boldsymbol{\tau}^* + \boldsymbol{\delta\tau} \geq \boldsymbol{0}$ and $\boldsymbol{\pi}^* + \boldsymbol{d\pi} \geq \boldsymbol{0}$. The dual optimum is therefore unique whenever no nonzero $\boldsymbol{d\tau}$ satisfies the first condition, which holds in particular when every owned link lies on some path containing no other owned link. Every dual optimum generates the same $(\boldsymbol{x}^*, \boldsymbol{y}^*)$, since Eq. (28) depends on $\boldsymbol{\tau}$ only through the path sums $T_k$ and Eq. (33) only through $\boldsymbol{\tau} - \boldsymbol{\pi}$.*

3. ***Two-sided outside options**. The outside path $k_i^0$ and the undeployed capacity $y_{m0}$ enter Eq. (28) and Eq. (33) as null alternatives of index 0 under the normalizations $\Phi_{i0} = 0$ and the reserve cost $c_{idle,m}$, contributing the terms 1 and $\exp(-\alpha_{S,m} c_{idle,m})$ in the denominators. They satisfy $x_{i0}^* = \exp(-\alpha_B u_i^*)$ and $y_{m0}^* = \exp\left(\alpha_{S,m}(-c_{idle,m} - v_m^*)\right)$, so the equilibrium utility of segment $i$ and the shadow price of operator $m$'s capacity are read directly off the unmatched flow and the reserve fleets. These are the structural counterparts of $\mu_{i0}$ and $\mu_{0j}$ in Choo and Siow (2006) and Galichon and Salanié (2022), and they anchor the identification of $\boldsymbol{\Phi}$ and $\boldsymbol{c}$ respectively.*

4. ***Preservation of dual structure under capacity rationing**. The flow capacity multiplier $\pi_a$ has no counterpart in Galichon and Salanié (2022). It nonetheless leaves the convex-duality structure intact: $\pi_a$ enters the seller response in Eq. (31) only through the affine link margin $\tau_a - c_a - \pi_a$, so it shifts the systematic index of the operator's logit allocation without altering its form. By complementary slackness $\pi_a^* > 0$ only where $y_a^* = C_a^{max}$, and a positive rent enters link $a$'s systematic index additively, strictly lowering its allocation while strictly raising that of operator $m$'s other links, holding the remaining multipliers fixed; at $\boldsymbol{\pi}^* = \boldsymbol{0}$ the pure margin-based logit share is recovered.*

***Proof.***

a. **Formulating the Lagrangian**. Relaxing the demand conservation constraints in Eq. (21) with multipliers $u_i$, the link market clearing constraints in Eq. (22) with $\tau_a \geq 0$, the flow capacity constraints in Eq. (23) with $\pi_a \geq 0$, and the operator capacity budget constraints in Eq. (24) with free multipliers $v_m$ gives Eq. (26), whose stationarity conditions in $x_{ik}$ and $y_a$ are Eq. (27) and Eq. (30).

b. **Deriving the dual conjugate**. $Z(\boldsymbol{x}, \boldsymbol{y})$ is strictly concave and separable across $\boldsymbol{x}$ and $\boldsymbol{y}$, so the two blocks may be maximized independently. For path flows $x_{ik}$, $\max_{x_{ik} \geq 0}\left\{(\Phi_{ik} - u_i - \sum_{a \in A_k \cap A^{own}} \tau_a) x_{ik} - \frac{1}{\alpha_B} x_{ik}(\ln x_{ik} - 1)\right\} = \frac{1}{\alpha_B}\exp(\alpha_B(\Phi_{ik} - u_i - \sum_{a \in A_k \cap A^{own}} \tau_a))$, yielding Eq. (28); at $k = k_i^0$ the optimum reduces to $x_{i0}^* = \exp(-\alpha_B u_i^*)$. For operator capacity, $\max_{y_a \geq 0}\left\{(\tau_a - c_a - \pi_a - v_m) y_a - \frac{1}{\alpha_{S,m}} y_a(\ln y_a - 1)\right\} = \frac{1}{\alpha_{S,m}}\exp\left(\alpha_{S,m}(\tau_a - c_a - \pi_a - v_m)\right)$, yielding Eq. (31); at the null alternative the expression reduces to $y_{m0}^* = \exp\left(-\alpha_{S,m}(c_{idle,m} + v_m^*)\right)$. Collecting the terms gives the Lagrangian dual $W$ in Eq. (35).

$$\min_{\boldsymbol{u}, \boldsymbol{\tau} \geq \boldsymbol{0}, \boldsymbol{\pi} \geq \boldsymbol{0}, \boldsymbol{v}} W = \frac{1}{\alpha_B} \sum_{i \in Q} \sum_{k \in K_i} \exp(\alpha_B(\Phi_{ik} - u_i - T_k))$$
$$+ \sum_{m \in M} \frac{1}{\alpha_{S,m}} \left( \sum_{a \in A_m} \exp\left(\alpha_{S,m}(\tau_a - c_a - \pi_a - v_m)\right) + \exp\left(\alpha_{S,m}(-c_{idle,m} - v_m)\right) \right)$$
$$+ \sum_{i \in Q} u_i D_i + \sum_{m \in M} \sum_{a \in A_m} \pi_a C_a^{max} + \sum_{m \in M} v_m F_m \tag{35}$$

Setting $\frac{\partial \mathcal{W}}{\partial u_i} = 0$ recovers Eq. (21) and gives Eq. (29); setting $\frac{\partial \mathcal{W}}{\partial v_m} = 0$ recovers Eq. (24) and gives Eq. (32). The null alternative contributes unit terms since $\Phi_{i0} = 0$ and the idle alternative carries margin $-c_{idle,m}$, contributing $\exp(-\alpha_{S,m} c_{idle,m})$. Substituting both closed forms back into Eq. (35) eliminates $\boldsymbol{u}$ and $\boldsymbol{v}$ and yields Eq. (34), with residual constants $\frac{D_i}{\alpha_B}$ and $\frac{F_m}{\alpha_{S,m}}$.

c. **The dual solution set**. $\mathcal{W}$ is a sum of logsums of affine functions plus a linear term, so it is convex. Along a segment between two minimizers $\mathcal{W}$ is constant, so each logsum has zero curvature in that direction. A logsum has zero curvature only if all its exponents shift by a common constant. Each logsum contains a fixed null alternative (the outside path or idle capacity), so that constant is zero. This gives $\sum_{a \in A_k} d\tau_a = 0$ for every path and $\boldsymbol{d\tau} = \boldsymbol{d\pi}$, and constancy of the remaining linear term gives $\sum_m \sum_a d\pi_a C_a^{max} = 0$. Any such direction leaves $\mathcal{W}$ unchanged.

d. **Strong duality and uniqueness**. The constraints are affine and $Z$ is concave, so strong duality holds if there is a feasible point with $\boldsymbol{x} > \boldsymbol{0}, \boldsymbol{y} > \boldsymbol{0}$. Take $y_a = \frac{1}{2} \frac{F_m C_a^{max}}{\sum_{a' \in A_m} C_{a'}^{max}}$ and $y_{m0} = \frac{1}{2}$. Put a small flow on every non-outside path and the remainder of $D_i$ on $k_i^0$; Eq. (22) then holds for a small enough flow. The feasible set is compact ($x_{ik} \leq D_i, y_a \leq F_m$) and $Z$ is continuous and strictly concave, so $(\boldsymbol{x}^*, \boldsymbol{y}^*)$ exists and is unique.

e. **Outside options and capacity rationing**. *Item 3.* These are the null-alternative cases of part (b): $u_i^* = -\frac{1}{\alpha_B} \ln x_{i0}^*$ and $v_m^* = -c_{idle,m} - \frac{1}{\alpha_{S,m}} \ln y_{m0}^*$. Since $\Phi_{i0} = 0$ and $c_{idle,m}$ fix the levels of $\boldsymbol{\Phi}$ and $v$, they anchor identification.
*Item 4.* Eq. (30) depends on $\pi_a$ only through the affine margin $\tau_a - c_a - \pi_a$, so Eq. (33) is a logit with $\pi_a$ entering the systematic index additively. Differentiating Eq. (33),
$$\frac{\partial \ln y_a^*}{\partial \pi_a} = -\alpha_{S,m}(1 - y_a^*/F_m) < 0, \qquad \frac{\partial \ln y_{a'}^*}{\partial \pi_a} = \alpha_{S,m} y_a^*/F_m > 0, \qquad a' \neq a$$
where $0 < y_a^*/F_m < 1$ because the denominator of Eq. (33) contains $\exp(-\alpha_{S,m} c_{idle,m}) > 0$. Complementary slackness on Eq. (23) gives $\pi_a^* > 0$ only where $y_a^* = C_a^{max}$, and at $\pi^* = 0$ the margin reduces to $\tau_a - c_a$. ■

**Corollary 1. The regularized core of P1**. *Let $\mathcal{C}$ denote the set of minimizers of Eq. (34) over $\{\boldsymbol{\tau} \geq \boldsymbol{0}, \boldsymbol{\pi} \geq \boldsymbol{0}\}$, with $\boldsymbol{u}$ and $\boldsymbol{v}$ recovered by Eqs. (29) and (32) and recentered as $\tilde{u}_i, \tilde{v}_m$. Then for every $(\boldsymbol{\tau}^*, \boldsymbol{\pi}^*) \in \mathcal{C}$:*

*(i)* *no traveler-operator coalition is blocking, in that*
$$\tilde{u}_i + T_k^* \geq \Phi_{ik}\ \forall i \in Q, k \in K_i, \qquad \tilde{v}_m + \pi_a^* \geq \tau_a^* - c_a, \forall a \in A_m, m \in M$$
*with slacks* $-\frac{1}{\alpha_B} \ln\left(\frac{x_{ik}^*}{D_i}\right) \geq 0$ *and* $-\frac{1}{\alpha_{S,m}} \ln\left(\frac{y_a^*}{F_m}\right) \geq 0$ *respectively;*

*(ii)* *payoffs exhaust the social surplus $\tilde{Z}^*$,*
$$\sum_{i \in Q} D_i \tilde{u}_i + \sum_{m \in M} F_m \tilde{v}_m + \sum_{m \in M} \sum_{a \in A_m} \pi_a^* C_a^{max} = \tilde{Z}^*;$$

*(iii)* *as $\alpha_B, \alpha_{S,m} \to \infty$ the slacks in (i) vanish on the coalitions that carry positive flow in the limit, and every limit point of $\mathcal{C}$ lies in the core characterized by Eqs. (5) – (6), extended by the capacity rents $\pi^*$, which have no counterpart there.*

***Proof***. (i) Rearranging Eq. (27) gives $\tilde{u}_i + \sum_{a \in A_k \cap A^{own}} \tau_a^* = \Phi_{ik} - \frac{1}{\alpha_B} \ln\left(\frac{x_{ik}^*}{D_i}\right)$, and the slack is nonnegative because $x_{ik}^* \leq D_i$ by Eq. (21); similarly from Eq. (30) with $y_a^* \leq F_m$ by Eq. (24). (ii) By strong duality $Z^* = W^*$; substituting Eqs. (29) and (32) into Eq. (35) and adding the constants of Eq. (20a) cancels the terms $\frac{1}{\alpha_B} \sum_i D_i \ln D_i$ and $\sum_m \frac{1}{\alpha_{S,m}} F_m \ln F_m$ together with $\sum_i \frac{D_i}{\alpha_B}$ and $\sum_m \frac{F_m}{\alpha_{S,m}}$. (iii) The inequalities in (i) pass to the limit, giving $\tilde{u}_i^\infty + \sum_{a \in A_k \cap A^{own}} \tau_a^\infty \geq \Phi_{ik}\ \forall\ k \in K_i$, the buyer half of the analogue of Eq. (6). Likewise, Eq.

(6) with the rent $\pi_a^\infty$ added. For the slacks, Eq. (29) writes $\tilde{u}_i$ as the logsum $\left(\frac{1}{\alpha_B}\right)\ln\sum_k \exp\left(\alpha_B(\Phi_{ik}-T_k)\right)$, so $\max_k(\Phi_{ik}-T_k)\le \tilde{u}_i \le \max_k(\Phi_{ik}-T_k)+\frac{\ln|K_i|}{\alpha_B}$, and hence $\tilde{u}_i-\max_k(\Phi_{ik}-T_k)\to 0$. The slack of path $k$ is $\tilde{u}_i-(\Phi_{ik}-T_k)$, which therefore converges to $\max_j\left(\Phi_{ij}-T_j^\infty\right)-\left(\Phi_{ik}-T_k^\infty\right)$, with zero precisely on the paths attaining the maximum and strictly positive elsewhere. The same argument applied to Eq. (32), whose logsum carries the idle term $\exp\left(-\alpha_{S,m}c_{idle,m}\right)$, gives the operator statement. ■

**Remark 1. Decentralization and flow-dependent link costs.** *The objective in Eq. (20) is the joint surplus of the traveler-operator coalitions, so the equilibrium of Eqs. (20) – (25) is the entropy-regularized system optimum. It is nonetheless supported by decentralized behavior: travelers choose paths to maximize their own perceived utility net of the cost transfers $\tau_a$ they pay, operators allocate capacity to maximize their own margin $\tau_a-c_a-\pi_a$ over their endowment $F_m$, and neither internalizes any externality directly. The market-clearing prices reconcile the two: by Theorem 1, the cost transfers that clear Eq. (22) make the selfish equilibrium coincide with the system optimum. Constraint (22) and its multiplier are therefore the mechanism of decentralization, and, absent flow-dependent link costs, the model is a first-best market rather than a centrally directed allocation. This is the network analogue of the role of cost transfers in Shapley and Shubik (1971), where the core is characterized by the dual variables in Eqs. (5) – (6), extended here to coalitions spanning multiple operators along a path. By Corollary 1, the dual solution set $\mathcal{C}$ is the entropy-regularized core, and it collapses to the core of Eqs. (5) – (6) in the deterministic limit.*

*The same mechanism accommodates flow-dependent link costs. Let $f_a=\sum_{i\in Q}\sum_{k\in K_i}\delta_{ak}x_{ik}$ and augment Eq. (20) with the total system cost $-\sum_{m\in M}\sum_{a\in A_m} f_a t_a(f_a,y_a)$, where $t_a$ is increasing in flow and decreasing in allocated capacity, such as the BPR function. Since $f_a^{n+1}/y_a^n$ is jointly convex in $(f_a,y_a)$ for $n\ge 0$, $Z$ remains strictly concave and the feasible set is unchanged, so strict concavity and uniqueness of the primal are preserved. Eqs. (28) and (33) hold only as a fixed point, so $\boldsymbol{u}$ and $\boldsymbol{v}$ can no longer be concentrated out and the box-constrained dual of Eq. (34) is unavailable. The traveler's generalized cost becomes $\tau_a+t_a(f_a,y_a)+f_a\frac{\partial t_a}{\partial f_a}$. Travelers do not internalize the third term by choice, and it cannot be carried by $\tau_a$, since complementary slackness on Eq. (22) forces $\tau_a^*=0$ wherever $f_a^*<y_a^*$. Supporting the system optimum therefore requires an explicit congestion toll alongside $\tau_a^*$. With that toll and as $\alpha_B\to\infty$, the equilibrium recovers Wardrop's second principle; without it, it is a stochastic user equilibrium at average cost $t_a$. Estimation under congestion would further require a bilevel inverse problem with equilibrium constraints (Chow et al., 2014; Bertsimas et al., 2015; Liu and Yin, 2026).*

**Remark 2. Economic interpretations.** *The model simultaneously outputs the traveler flows, operator capacity designs, and first best pricing needed to clear the market to attain a social optimum. The model further provides economic interpretations from a number of its other outputs.*

*Value of $\tilde{v}_m^*$. Computing the value of $\tilde{v}_m^*$ as a function of available capacity $F_m$, by re-solving P1 at each $F_m$, traces a derived demand curve for operator m's capacity. Because $\tilde{v}_m^*$ is the marginal social surplus of an additional unit, it supports optimal capacity sizing: invest until $\tilde{v}_m^*$ equals the per-unit acquisition cost. It also supports subsidy targeting, since a dollar of subsidy buys the most social surplus at the operator with the highest $\tilde{v}_m^*$.*

*Market entry/exit. By Eqs. (32) and (33), the reserve fraction is $\frac{y_{m0}^*}{F_m}=\exp\left(-\alpha_{S,m}\left(c_{idle,m}+\tilde{v}_m^*\right)\right)$, a continuous participation margin that depends on $F_m$ only through the equilibrium. An operator with $\tilde{v}_m^*\to -c_{idle,m}$ holds its entire endowment in reserve and is effectively exiting the market, so entry and exit become threshold conditions on $\tilde{v}_m^*$.*

*Network value via logsums. The recentered duals $\tilde{u}_i^*$ and $\tilde{v}_m^*$ are the buyer-side and seller-side logsums, and differences in them quantify the value of network and market elements (e.g. hubs, corridors, presence/absence of a traveler segment or operator) to travelers and operators.*

*Cross-operator externalities. The partial derivatives* $\frac{\partial \tilde{u}_i^*}{\partial F_m}$ *and* $\frac{\partial \tilde{v}_{m'}^*}{\partial F_m}$ *can be computed, which informs on the impact of an operator* $m$*'s change in their endowment* $F_m$ *on different traveler segments* $i \in Q$ *and other operators* $m' \neq m$.

The primal problem is an entropy-regularized convex program with linear constraints. It exhibits strict concavity, so it is guaranteed to have a unique global optimum $(\boldsymbol{x}^*, \boldsymbol{y}^*)$. The dual function $\mathcal{W}$ is continuously differentiable. With the box constrained dual, we can solve this iteratively using dual quasi-Newton methods, i.e. L-BFGS-B (Byrd et al., 1995).

**Remark 3. Convergence criterion**. *Let* $\boldsymbol{z} = (\boldsymbol{\tau}, \boldsymbol{\pi})$. *Since Eq. (34) minimizes a convex, continuously differentiable function over the nonnegative orthant,* $z^*$ *is optimal iff it satisfies the first order conditions in Eq. (36).*

$$0 \leq z_j \perp \frac{\partial \mathcal{W}}{\partial z_j} \geq 0, \qquad \forall j \tag{36}$$

*Substituting the gradient expressions from Eq. (34) leads to the expressions in Eq. (37).*

$$\frac{\partial \mathcal{W}}{\partial \tau_a} = y_a - f_a, \qquad \frac{\partial \mathcal{W}}{\partial \pi_a} = C_a^{max} - y_a \tag{37}$$

*To measure how far an iterate is from satisfying Eq. (36), we can use a gap-type function. However, both sets of dual variables can be set to zero, which can lead to degenerate solutions if applying a more conventional complementarity gap like in Boyce et al. (2004), where the assignment iterates load the full demand and are always feasible. Dual iterates are not always feasible. Instead, we consider a natural residual gap function (see Facchinei and Pang, 2003), which measures the largest complementarity violation. A natural residual is defined as* $NR(a,b) = \min(a,b)$ *where* $\min(a,b) = 0$ *iff* $a \geq 0, b \geq 0$, *and* $ab = 0$. *Applied component-wise to Eq. (36) and aggregated in the sup norm, this gives a scalar* $\varrho$ *that vanishes at the optimum, defined in Eq. (38).*

$$\varrho^{(t)} = \frac{1}{\max_a f_a^{(t)}} \max_{m \in M, a \in A_m} \left\{ \left| \min\left(\tau_a^{(t)}, y_a^{(t)} - f_a^{(t)}\right) \right|, \left| \min\left(\pi_a^{(t)}, C_a^{max} - y_a^{(t)}\right) \right| \right\} \tag{38}$$

*Eq. (38) shows the largest violation of market clearing on any link, expressed as a fraction of the busiest link's load. Since* $\varrho^{(t)}$ *is not the quantity being minimized, it need not decrease monotonically; excursions occur when a link enters or leaves the active set of Eq. (23).*

**Algorithm 1.** L-BFGS-B for dual solution of the path-flow formulation of the aggregate network assignment game P1

***Input.*** Network $A_0, \{A_m\}_{m \in M}$, path sets $\{K_i\}_{i \in Q}$, and incidence $\delta_{ak}$; demands $D_i$; endowments $F_m$; caps $C_a^{max}$; costs $c_a, c_{idle,m}$; surpluses $\Phi_{ik}$ with $\Phi_{i0} = 0$; scales $\alpha_B, \alpha_{S,m}$; tolerance $\epsilon$.
***Output.*** $(\boldsymbol{\tau}^*, \boldsymbol{\pi}^*)$, and the primal $(\boldsymbol{x}^*, \boldsymbol{y}^*)$ with $(\boldsymbol{u}^*, \boldsymbol{v}^*)$.
1. *Initialize.* $\boldsymbol{\tau}^{(0)} \leftarrow \boldsymbol{0}, \boldsymbol{\pi}^{(0)} \leftarrow \boldsymbol{0}$.
2. Minimize Eq. (34) over the box $\boldsymbol{\tau} \geq \boldsymbol{0}, \boldsymbol{\pi} \geq \boldsymbol{0}$ by L-BFGS-B. At each iterate $(\boldsymbol{\tau}, \boldsymbol{\pi})^{(t)}$ (index $t$ dropped for convenience below):

   a. *Buyer loading*. For each $i \in Q$: path prices $T_k$; inclusive value $\tilde{u}_i = \ln(1 + \sum_{k \in K_i \setminus \{k_i^0\}} \exp(\alpha_B(\Phi_{ik} - T_k))$; then $x_{ik} = D_i \exp(\alpha_B(\Phi_{ik} - T_k - \tilde{u}_i))$ and $x_{i0} = D_i \exp(-\alpha_B \tilde{u}_i)$ by Eq. (28). Link loads $f_a = \sum_{i \in Q} \sum_{k \in K_i} \delta_{ak} x_{ik}$.

   b. *Seller loading*. For each $m \in M$: inclusive value $\tilde{v}_m = \ln\left(\exp(-\alpha_{S,m} c_{idle,m}) + \sum_{a \in A_m} \exp\left(\alpha_{S,m}(\tau_a - c_a - \pi_a)\right)\right)$; then $y_a = F_m \exp\left(\alpha_{S,m}(\tau_a - c_a - \pi_a - \tilde{v}_m)\right)$ and $y_{m0} = F_m \exp\left(-\alpha_{S,m}(c_{idle,m} + \tilde{v}_m)\right)$ by Eq. (33).

c. *Update objective value in Eq. (34) and gradient* $\frac{\partial w}{\partial \tau_a} = y_a - f_a, \frac{\partial w}{\partial \pi_a} = C_a^{max} - y_a$. A search direction is formed from a limited-memory inverse-Hessian approximation built from successive gradient differences, projects onto the bounds, and selects the step by line search; the iterate is accepted when Eq. (34) decreases.

d. *Stop* when $\varrho \le \epsilon$, where $\varrho$ is defined in Eq. (38).

3. Recover $u_i^*$ and $v_m^*$ from Eqs. (29) and (32), with $(\boldsymbol{x}^*, \boldsymbol{y}^*)$ from step 2(a)-(b).

In real world systems, however, the prices are already in place as observed $\boldsymbol{\tau}^{\mathbf{0}}$, which may not be optimal. A method is needed to estimate the parameters of the model such that the observed tolls and their effects on the flows make sense, so that an optimal set of tolls can then be derived from it. Real world systems typically lack available path flow data, with link count data more prevalent (Willumsen, 1978; Yang et al., 1992; Tebaldi and West, 1998). In addition, the path-flow formulation requires path set generation, which further prohibits the application of the models to scale. Finally, the path-based logit structure requires that every path has some positive flow, which may be unrealistic.

### 3.2. PURC-based formulation with service-level capacity of aggregate network assignment game

Borrowing from the method of Yang and Chow (2026), we convert P1 into P2, a PURC-based formulation (see Fosgerau et al., 2022) for which we can establish an inverse optimization problem. The primary trade-offs of this alternate formulation to the path-based aggregate assignment game include:

- *Pros*: Not having to enumerate paths, can estimate from link count data, admits zero flows on links, and handles path overlap
- *Cons*: Path flows are not explicitly identified despite the underlying path-based assignment mechanism, and link features need to be additive.

P2 also generalizes how operators deploy capacity. In P1 an operator allocates its endowment link by link. In many service systems, such as public transit, capacity is deployed on lines instead: a train provides the same seats on every link it traverses, so the capacity of a link is the sum over the lines using it. P2 therefore lets each operator allocate capacity to a set of services, each a fixed set of its links. Link-level allocation is the special case in which every service is a single link, so systems that mix the two are covered by the same formulation.

Let

$V$: set of nodes, where $o_i \subseteq V$ is the set of origins of segment $i \in Q$ with destination $s_i \in V$, where $|o_i| = 1$ is an OD-based segment as a special case,

$t(a), h(a) \in V$: tail and head nodes of link $a \in A$,

$D_{in}$: demand of segment $i$ originating at $n$ in $o_i$, with $\sum_{n \in o_i} D_{in} = D_i$,

$b_{in}$: net supply indicator for segment $i \in Q$ at node $n \in V$, equal to $D_{in}/D_i$ if $n \in o_i$, $-1$ if $n = s_i$, and 0 otherwise,

$a_{in}^0$: the outside link from $n \in o_i$ to $s_i$,

$\varphi_{ia}$: link-level surplus rate accruing to segment $i \in Q$ from using link $a \in A$, normalized to $\varphi_{i,a_{in}^0} = 0$ on the outside link,

$\ell_a > 0$: length weight of link $a \in A$,

$f_{ia} \ge 0$: flow of segment $i \in Q$ on link $a \in A$, where $f_a = \sum_{i \in Q} f_{ia}$,

$u_{in}$: node potential of segment $i \in Q$ at node $n \in V$, normalized to $u_{i,s_i} = 0$, this replaces the $u_i$ in P1,

$\psi_{ia}$: reduced link margin, where $\psi_{ia} = \varphi_{ia} - \tau_a - \left(u_{i,t(a)} - u_{i,h(a)}\right)$,

$R_m$: set of services operated by $m \in M$, with $R = \bigcup_{m \in M} R_m$. A service $r \in R_m$ is a fixed set of links in $A_m$ such as a train pattern from an origin station to a terminal, or a single link,

$\delta_{ar} \in \{0,1\}$: link-service incidence, equal to 1 if service $r$ uses link $a \in A_m$, and 0 otherwise,

$Y_r \ge 0$: capacity deployed on service $r \in R_m$. The capacity of link $a \in A_m$ is $y_a = \sum_{r \in R_m} \delta_{ar} Y_r$,

$c_r \equiv \sum_{a \in A_m} \delta_{ar} c_a$: unit cost of service $r$, since a unit of capacity on $r$ occupies every link it uses,

$\omega_r \equiv \sum_{a\in A_m} \delta_{ar}(\tau_a - c_a - \pi_a)$: service margin
$\tau_a^0 \geq 0$: observed cost transfers that may not be the optimal cost transfers,
$\sigma_a \geq 0\ \forall a \in A^{own},\ \sigma_a \equiv 0\ \forall a \in A_0$: the rationing price under non-optimal tolls $\tau_a^0$,
$\Delta_a = \tau_a^0 - \tau_a^*$: pricing wedge between the observed fare and the recovered clearing price $\tau_a^*$.

**Assumption 2** (link-additive surplus and service coverage). *The path surplus of P1 decomposes as* $\Phi_{ik} = \sum_{a\in A_k} \varphi_{ia}$ *for all* $i \in Q, k \in K_i$, *with* $\varphi_{i,a_{in}^0} = 0$ *preserving the normalization* $\Phi_{i0} = 0$. *No directed cycle* $\xi \subseteq A$ *satisfies* $\sum_{a\in\xi} \varphi_{ia} > 0$ *for any* $i \in Q$. *Every owned link* $a \in A_m$ *lies on at least one service in* $R_m$, *and each origin* $n \in o_i$ *has an outside link* $a_{in}^0 \in A_0$ *to* $s_i$.

P2:

$$\max_{f,Y,y_{m0}} Z' = \sum_{i\in Q}\sum_{a\in A} \varphi_{ia} f_{ia} - \sum_{m\in M}\sum_{r\in R_m} c_r Y_r - \sum_{m\in M} c_{idle,m} y_{m0} - \frac{1}{\alpha_B}\sum_{i\in Q}\sum_{a\in A}\frac{\ell_a f_{ia}^2}{2D_i} - \sum_{m\in M}\frac{1}{\alpha_{S,m}}\left(\sum_{r\in R_m} Y_r(\ln Y_r - 1) + y_{m0}(\ln y_{m0} - 1)\right) \tag{39}$$

s.t.

Buyer flow conservation:

$$\sum_{a:t(a)=n} f_{ia} - \sum_{a:h(a)=n} f_{ia} = D_i b_{in}, \qquad \forall i \in Q, n \in V \qquad (u_{in}) \tag{40}$$

Link market clearing:

$$\sum_{i\in Q} f_{ia} \leq \sum_{r\in R_m} \delta_{ar} Y_r, \qquad \forall a \in A_m, m \in M \qquad (\tau_a) \tag{41}$$

Link flow capacity:

$$\sum_{r\in R_m} \delta_{ar} Y_r \leq C_a^{max}, \qquad \forall a \in A_m, m \in M \qquad (\pi_a) \tag{42}$$

Budget constraint:

$$\sum_{r\in R_m} Y_r + y_{m0} = F_m, \qquad \forall m \in M \qquad (v_m) \tag{43}$$

Non-negativity:

$$f_{ia} \geq 0, \qquad \forall i \in Q, a \in A \tag{44}$$

$$Y_r \geq 0, \qquad \forall r \in R_m, m \in M, \qquad y_{m0} \geq 0, \qquad \forall m \in M \tag{45}$$

For the perturbation function of the perturbed utility route choice model, we assume a quadratic function $H_a(\zeta) = \frac{\ell_a}{2}\zeta^2$ which can obtain zero flows (Fosgerau et al., 2022). A link-level operator can be modeled with ($\delta_{ar} = 1$ iff $r = a$) and $Y_r = y_a$, so a market with both link-level and line-level operators can be captured. Note that outside links carry no market clearing or capacity constraint. Due to the buyer quadratic and seller entropy term, P2 is a strictly concave nonlinear program with linear constraints. As such, it is globally solvable with a unique solution. Consider the following first order conditions for the buyer block. Similar to Eq. (20), Eq. (39) is missing a constant term (only on the seller side) for it to reflect the social surplus. The following modified function in Eq. (39a) is used to compute it.

$$\tilde{Z}' = Z'^* + \sum_{m\in M}\frac{1}{\alpha_{S,m}}(F_m \ln F_m - F_m) \tag{39a}$$

Due to differences in regularizers, the social surplus values of P1 and P2 are not equivalent, so comparisons between their social surpluses are meaningless. However, the gross surplus is comparable. We define the gross surplus $G(\boldsymbol{\tau})$ as Eq. (46).

$$G(\boldsymbol{\tau}) = \sum_{i\in Q}\sum_{a\in A}\varphi_{ia}f_{ia} - \sum_{m\in M}\sum_{r\in R_m} c_r Y_r \tag{46}$$

Let the Lagrangian be defined as $\mathcal{L}'$. Then taking $\frac{\partial \mathcal{L}'}{\partial f_{ia}}$ with complementarity results in Eq. (47).

$$0 \le f_{ia}^* \perp \frac{\ell_a}{\alpha_B}\frac{f_{ia}^*}{D_i} - \varphi_{ia} + \tau_a^* + u_{i,t(a)}^* - u_{i,h(a)}^* \ge 0 \tag{47}$$

Eq. (47) leads to the following link flows in Eq. (48), where they are set to either the right-hand term set to zero or to zero flow.

$$f_{ia}^* = \frac{\alpha_B D_i}{\ell_a}\left(\varphi_{ia} - \tau_a^* - \left(u_{i,t(a)}^* - u_{i,h(a)}^*\right)\right)_+ \tag{48}$$

Rearranging terms in Eq. (47) in terms of the tail node potential gives $u_{i,t(a)} \ge \varphi_{ia} - \tau_a - \frac{\ell_a}{\alpha_B}\frac{f_{ia}^*}{D_i} + u_{i,h(a)}$ on all $a$ and on the flow-reachable subnetwork, which can then be written as Eq. (49).

$$u_{in}^* = \max_{a:t(a)=n}\left\{\varphi_{ia} - \tau_a^* - \frac{\ell_a}{\alpha_B}\frac{f_{ia}^*}{D_i} + u_{i,h(a)}^*\right\}, \qquad u_{i,s_i}^* = 0 \tag{49}$$

**Remark 4**. **Accessibility measure**. *The node potential at the origin $u_{in}^*, n \in o_i$, is the shadow price of flow conservation at origin $n$. It is obtained without enumerating paths. Because the perturbation in Eq. (39) is scaled by $D_i$, $u_{in}^*$ is already a per-traveler quantity and needs no recentering by $\frac{1}{\alpha_B}\ln D_i$. Its counterpart in P1 is therefore $\tilde{u}_i^*$ rather than $u_i^*$. The segment-level measure is the demand-weighted origin potential $\sum_{n\in o_i} b_{in}u_{in}^*$. When the outside link $a_{in}^0$ carries flow, the outside link relation gives $u_{in}^* = -\frac{\ell_{a_{in}^0}}{\alpha_B}\frac{f_{i,a_{in}^0}^*}{D_i} < 0$. As in Eq. (29), the traveler's payoff is then read directly off the unmatched flow, which preserves the two-sided outside-option structure. When the outside link carries no flow, which the quadratic perturbation permits, complementarity gives $u_{in} \ge 0$ instead. The measure is therefore not numerically comparable with $\tilde{u}_i^*$ in P1, which is a logsum with the outside option at index zero and so satisfies $\tilde{u}_i^* \ge 0$. Within either formulation, differences across scenarios are the meaningful accessibility comparison.*

Stationarity in $Y_r$ gives the service-level counterpart of Eq. (33) in Eq. (49a).

$$Y_r^* = F_m\frac{\exp\left(\alpha_{S,m}\omega_r^*\right)}{\exp\left(-\alpha_{S,m}c_{idle,m}\right) + \sum_{r'\in R_m}\exp\left(\alpha_{S,m}\omega_{r'}^*\right)} \tag{49a}$$

with $y_{m0}^*$ and $v_m^*$ given by Eqs. (31) – (32). The operator ranks services by their whole-service margin, so a link receives the capacity of every service that uses it, whatever its own share of the fare. With single-link services, Eq. (49a) is Eq. (33).

Similar to Eq. (34), the corresponding dual problem in Eq. (50) can concentrate out the closed form $v_m$ but now needs to include $u_{in}$ as one of the decision variables, where $\psi_{ia} = \varphi_{ia} - \tau_a - \left(u_{i,t(a)} - u_{i,h(a)}\right)$.

$$\min_{u,\tau\geq 0,\pi\geq 0} \mathcal{W}' = \sum_{i\in Q}\sum_{a\in A}\frac{\alpha_B D_i}{2\ell_a}[\psi_{ia}]_+^2 + \sum_{i\in Q}\sum_{n\in o_i} D_i b_{in} u_{in}$$
$$+ \sum_{m\in M}\frac{F_m}{\alpha_{S,m}}\left(\ln\left(\exp(-\alpha_{S,m}c_{idle,m}) + \sum_{r\in R_m}\exp(\alpha_{S,m}\omega_r)\right) - \ln F_m + 1\right)$$
$$+ \sum_{m\in M}\sum_{a\in A_m}\pi_a C_a^{max} \quad (50)$$

**Theorem 2**. **Duality and uniqueness of the PURC-based formulation.** *Let P2 in Eqs. (39) – (45) satisfy Assumptions 1 and 2. Then: (i) $Z'$ is strictly concave and the primal optimum $(\boldsymbol{f}^*, \boldsymbol{Y}^*)$ is unique, with no cyclic flow; (ii) the dual in Eq. (50) is jointly convex on $\{\boldsymbol{u} \in \mathbb{R}^{|Q||V|}, \boldsymbol{\tau} \geq \boldsymbol{0}, \boldsymbol{\pi} \geq \boldsymbol{0}\}$ and strong duality holds; (iii) any dual minimizer generates $(\boldsymbol{f}^*, \boldsymbol{Y}^*)$ through Eqs. (48) and (49a), with $u^*$ unique up to the normalization $u_{i,s_i}^* = 0$ and on the flow-reachable subnetwork; (iv) the outside link $a_{in}^0$ and the idle capacity $y_{m0}$ retain their roles as the two null alternatives, with the origin potentials $u_{in}^*, n \in o_i$, and $v_m^*$ anchoring identification of $\boldsymbol{\varphi}$ and $\boldsymbol{c}$ respectively.*

***Proof.***
The proof follows that of Theorem 1, with (b) Deriving the dual conjugate replaced and a new part (f) on no cyclic flows.

b) **Deriving the dual conjugate**. $Z'$ is separable in $\boldsymbol{f}$ and $\boldsymbol{Y}$. For each cell, $\max_{f_{ia}\geq 0}\left\{\psi_{ia}f_{ia} - \frac{\ell_a}{\alpha_B}\frac{f_{ia}^2}{2D_i}\right\} = \frac{\alpha_B D_i}{2\ell_a}[\psi_{ia}]_+^2$, attained at Eq. (48). For each service, $\max_{Y_r\geq 0}\left\{(\omega_r - v_m)Y_r - \frac{1}{\alpha_{S,m}}Y_r(\ln Y_r - 1)\right\} = \frac{1}{\alpha_{S,m}}\exp\left(\alpha_{S,m}(\omega_r - v_m)\right)$, because $\omega_r$ collects the multipliers of Eqs. (41) – (42) on the links of $r$. Imposing Eq. (43) gives Eq. (49a), and concentrating $v_m$ out gives the seller term of Eq. (50). Relaxing Eq. (40) contributes $\sum_{i\in Q}\sum_{n\in o_i} D_i b_{in} u_{in}$ under $u_{i,s_i} = 0$. $u_{in}$ has no closed form and remains a decision variable.
f) **No cyclic flow**. Suppose segment $i$ has positive flow on every link of a directed cycle. Subtracting the cycle's smallest link flow from each of its links keeps Eqs. (40) – (41) and nonnegativity at the same $\boldsymbol{Y}^*$. By Assumption 2 this does not decrease the linear term of $Z'$, and it strictly decreases the quadratic perturbation, contradicting optimality. ■

Unlike Theorem 1(2), dual uniqueness here is conditional on the active set. Two dual minimizers can differ only along directions $(du, d\tau, d\pi)$ that satisfy three conditions: $d\tau_a + du_{i,t(a)} - du_{i,h(a)} = 0$ on cells with $f_{ia}^* > 0$; $\sum_a \delta_{ar}(d\tau_a - d\pi_a) = 0$ for every service $r$; and $\sum_{a\in A_m} d\pi_a C_a^{max} = 0$. The squared hinge in Eq. (50) is flat wherever $\psi_{ia} < 0$, so inactive cells add no restriction. Individual link prices are separated by services that share some links but not others, and by buyer paths that use part of a service. With single-link services, the second condition is $d\tau_a = d\pi_a$, as in Theorem 1.

Similar to P1, P2 can be solved through the dual problem, which is convex and continuously differentiable. Algorithm 1 (L-BFGS-B) extends directly to Algorithm 2, with $u$ entering as free variables alongside the box constrained $(\boldsymbol{\tau}, \boldsymbol{\pi})$ and Eq. (48) replacing the buyer loading step. For the convergence criteria, Eq. (38) is modified to Eq. (38a) to further account for the node potentials.

$$\varrho^{(t)} = \frac{1}{\max_a f_a^{(t)}}\max\left\{\max_{i\in Q,n\in V}\left|\eta_{in}^{(t)}\right|, \max_{m\in M,a\in A_m}\left\{\left|\min\left(\tau_a^{(t)}, y_a^{(t)} - f_a^{(t)}\right)\right|, \left|\min\left(\pi_a^{(t)}, C_a^{max} - y_a^{(t)}\right)\right|\right\}\right\},$$
$$y_a^{(t)} = \sum_{r\in R_m}\delta_{ar}Y_r^{(t)} \quad (38a)$$

**Algorithm 2.** L-BFGS-B for dual solution of P2

***Input.*** Network $(V, A)$ with tail and head maps $t(a), h(a)$, operator link sets $\{A_m\}_{m\in M}$, and outside links $a_{in}^0$; service sets $\{R_m\}_{m\in M}$ with link-service incidence $\delta_{ar}$. For a link-level operator, each service is a single link; demands $D_i$ and net supply indicators $b_{in}$; endowments $F_m$; caps $C_a^{max}$; costs $c_a$ and $c_{idle,m}$; link surpluses $\varphi_{ia}$; length weights $\ell_a$; scales $\alpha_B, \{\alpha_{S,m}\}_{m\in M}$; tolerance $\epsilon$.

***Output.*** $(\boldsymbol{u}^*, \boldsymbol{v}^*, \boldsymbol{\tau}^*, \boldsymbol{\pi}^*)$, and the primal $(\boldsymbol{f}^*, \boldsymbol{Y}^*)$ with link capacities $y_a^* = \sum_r \delta_{ar} Y_r^*$.

1. *Initialize.* $\boldsymbol{\tau}^{(0)} \leftarrow \boldsymbol{0}, \boldsymbol{\pi}^{(0)} \leftarrow \boldsymbol{0}$; set $u_{in}^{(0)}$ by a shortest path pass on $-\varphi_{ia} + \ell_a/\alpha_B$ from $s_i$, with $u_{i,s_i} = 0$.
2. Minimize Eq. (50) by L-BFGS-B over $\boldsymbol{u}$ free and the box $\boldsymbol{\tau} \geq \boldsymbol{0}, \boldsymbol{\pi} \geq \boldsymbol{0}$. Enforce $u_{i,s_i} = 0$ by eliminating those variables. At each iterate $(\boldsymbol{u}, \boldsymbol{\tau}, \boldsymbol{\pi})^{(t)}$ (index $t$ dropped below):
   a. *Buyer loading.* For each $i \in Q$ and $a \in A$: form the reduced margin $\psi_{ia} = \varphi_{ia} - \tau_a - \left(u_{i,t(a)} - u_{i,h(a)}\right)$. Then $f_{ia} = \left(\frac{\alpha_B D_i}{\ell_a}\right)[\psi_{ia}]_+$ by Eq. (48). Compute link loads $f_a = \sum_{i\in Q} f_{ia}$ and node residuals $\eta_{in} = \sum_{a:t(a)=n} f_{ia} - \sum_{a:h(a)=n} f_{ia} - D_i b_{in}$.
   b. *Seller loading.* Form service margin $\omega_r = \sum_{a\in A_m} \delta_{ar}(\tau_a - c_a - \pi_a)$. For each $m \in M$ : $\tilde{v}_m = \frac{1}{\alpha_{S,m}} \ln\left(\exp(-\alpha_{S,m} c_{idle,m}) + \sum_{r\in R_m} \exp(\alpha_{S,m}\omega_r)\right)$. Then service capacities $Y_r = F_m \exp\left(\alpha_{S,m}\omega_r - \tilde{v}_m\right)$ and $y_{m0} = F_m \exp\left(-\alpha_{S,m}\left(c_{idle,m} + \tilde{v}_m\right)\right)$ by Eq. (49a); and link capacities $y_a = \sum_{r\in R_m} \delta_{ar} Y_r$.
   c. *Update objective value in Eq. (50) and gradient:* $\frac{\partial W'}{\partial u_{in}} = -\eta_{in}, \frac{\partial W'}{\partial \tau_a} = y_a - f_a, \frac{\partial W'}{\partial \pi_a} = C_a^{max} - y_a$. A search direction is formed from a limited-memory inverse-Hessian approximation built from successive gradient differences and projected onto the bounds. A line search selects the step, and the iterate is accepted when Eq. (50) decreases.
   d. *Stop* when $\varrho \leq \epsilon$, where $\varrho$ is defined as Eq. (38a).
3. Recover $v_m^*$ from the $\tilde{v}_m^*$ in step 2(b), with $(\boldsymbol{f}^*, \boldsymbol{Y}^*)$ from step 2(a)-(b). The equilibrium utility of segment $i$ at origin $n \in o_i$ is $u_{in}^*$. Equivalently, it is $-\left(\frac{\ell_{a_{in}^0}}{\alpha_B}\right)\left(\frac{f_{i,a_{in}^0}^*}{D_i}\right)$ from the outside link.

While Algorithm 2 outputs the $(\boldsymbol{f}^*, \boldsymbol{Y}^*, \boldsymbol{\tau}^*, \boldsymbol{\pi}^*, \boldsymbol{u}^*, \boldsymbol{v}^*)$, it is solving the first best pricing problem in P2. In estimating parameters, an observed market's prices $\boldsymbol{\tau}^{\boldsymbol{0}}$ may not equal the $\boldsymbol{\tau}^*$. A variant of Algorithm 2 is introduced to find the rest of the primal and dual variables given a fixed price $\boldsymbol{\tau}^{\boldsymbol{0}}$ for an uncleared market. At an administered fare $\boldsymbol{\tau}^{\boldsymbol{0}}$, the capacity constraint must be carried explicitly rather than dualized. We define a new variable $\sigma_a \geq 0$ where travelers face the effective price $\tau_a^0 + \sigma_a$, i.e. the posted fare plus the shadow price of scarce capacity. This $\sigma_a$ is a rationing price and not a revenue that the operator would receive. Algorithm 2' is presented to obtain that solution.

**Algorithm 2'.** Solution to P2 with fixed prices $\boldsymbol{\tau}^{\boldsymbol{0}}$

***Input.*** Network $(V, A), A = \{A_0, \{A_m\}_{m\in M}\}$; node-link structure $t(a), h(a), b_{in}$; service sets $\{R_m\}_{m\in M}$ with link-service incidence $\delta_{ar}$; segment $i$ demands $D_i$; endowments $F_m$; caps $C_a^{max}$; costs $c_a, c_{idle,m}$; observed prices $\boldsymbol{\tau}^{\boldsymbol{0}}$; link surpluses $\varphi_{ia}$; length weights $\ell_a$; scales $\alpha_B, \{\alpha_{S,m}\}_{m\in M}$; tolerance $\epsilon$; averaging weight function $\gamma_t = t^{-0.6}$.

***Output.*** $(\boldsymbol{u}^\dagger, \boldsymbol{v}^\dagger, \boldsymbol{\sigma}^\dagger, \boldsymbol{\pi}^\dagger)$, and the primal $(\boldsymbol{f}^\dagger, \boldsymbol{Y}^\dagger)$ with link capacities $y_a^\dagger = \sum_r \delta_{ar} Y_r^\dagger$; load factors $\boldsymbol{L}^\dagger$.

1. *Initialize.* Set $t = 1$; $\boldsymbol{Y}^{(1)}$ from seller step at $\boldsymbol{L} \equiv 1$ and set $y_a^{(1)} = \sum_r \delta_{ar} Y_r^{(1)}$. Initialize $\boldsymbol{u}$ by a shortest path pass on $-(\varphi_{ia} - \tau_a^0)$ from each sink $s_i$.
2. *Buyer step.* Given $\boldsymbol{y}^{(t)}$, minimize over $(\boldsymbol{u}, \boldsymbol{\sigma} \geq 0)$

$$\sum_{i\in Q}\sum_{a\in A} \frac{\alpha_B D_i}{2\ell_a}[\psi_{ia}]_+^2 + \sum_{i\in Q}\sum_{n\in V} D_i b_{in} u_{in} + \boldsymbol{\sigma}^T \boldsymbol{y}^{(t)}$$

   where $\psi_{ia} = \varphi_{ia} - \tau_a^0 - \sigma_a - (u_{i,t(a)} - u_{i,h(a)})$ and $f_{ia} = \frac{\alpha_B D_i}{\ell_a}[\psi_{ia}]_+$. The gradients are $\frac{\partial}{\partial u_{in}} = -\eta_{in}$ and $\frac{\partial}{\partial \sigma_a} = y_a^{(t)} - f_a$.

3. *Seller step*. From the link flows $f_a^{(t)} = \sum_{i\in Q} f_{ia}^{(t)}$ from step 2, form load factors $L_a^{(t)} = \min\left(1, \frac{f_a^{(t)}}{y_a^{(t)}}\right)$. Then solve per operator $m \in M$ as a best response given perceived marginal revenue $\tau_a^0 L_a^{(t)}$:

$$\max_{\boldsymbol{Y}\geq\boldsymbol{0}, y_{m0}\geq 0} S_m = \sum_{r\in R_m}\sum_{a\in A_m} \delta_{ar}\left(\tau_a^0 L_a^{(t)} - c_a\right) Y_r - c_{idle,m} y_{m0} - \frac{1}{\alpha_{S,m}}\left[\sum_{r\in R_m} Y_r(\ln Y_r - 1) + y_{m0}(\ln y_{m0} - 1)\right]$$

s.t.

Eqs. (42) – (43)

returning auxiliary values $\tilde{Y}_r^{(t)} = F_m \exp\left(\alpha_{S,m}\left(\sum_{a\in A_m} \delta_{ar}\left(\tau_a^0 L_a^{(t)} - c_a\right) - \sum_a \delta_{ar}\pi_a - \tilde{v}_m\right)\right)$, $\tilde{v}_m = \frac{1}{\alpha_{S,m}} \ln\left(\exp(-\alpha_{S,m} c_{idle,m}) + \sum_{r\in R_m} \exp\left(\alpha_{S,m}\left(\sum_{a\in A_m} \delta_{ar}\left(\tau_a^0 L_a^{(t)} - c_a\right) - \sum_a \delta_{ar}\pi_a\right)\right)\right)$, $v_m^{(t)} = \tilde{v}_m - \frac{1}{\alpha_{S,m}} \ln F_m$, and the capacity rent $\boldsymbol{\pi}^{(t)} \geq 0$. The rents minimize the smooth convex dual. If no cap binds at $\boldsymbol{\pi} = 0$, then $\boldsymbol{\pi}^{(t)} = 0$. With single-link services, capped links are fixed at $C_a^{max}$ and the remaining budget is shared by Eq. (33).

4. *Average and test*. Set $\boldsymbol{Y}^{(t+1)} = (1 - \gamma_t)\boldsymbol{Y}^{(t)} + \gamma_t \tilde{\boldsymbol{Y}}^{(t)}$ and $y_a^{(t+1)} = \sum_r \delta_{ar} Y_r^{(t+1)}$. If $\varrho^{(t)} \leq \epsilon$ (from Eq. (38a), with $\tau_a^{(t)}$ replaced by $\tau_a^0 + \sigma_a^{(t)}$), stop; otherwise set $t := t + 1$ and return to step 2.

† is used to denote optimal values under the fixed price variant of P2.

Existence of an equilibrium under Algorithm 2' is shown in Theorem 3.

**Theorem 3. Existence of a fixed price equilibrium**. *Let P2 satisfy Assumptions 1 and 2. For any administered fare $\boldsymbol{\tau^0} \geq 0$, the map defined by steps 2-3 of Algorithm 2' admits a fixed point $\boldsymbol{Y}^*$. At any such fixed point, travelers and operators are simultaneously best responding at $\boldsymbol{\tau^0}$. Then $\boldsymbol{Y}^*$, with link capacities $y_a^* = \sum_r \delta_{ar} Y_r^*$ and the associated $(\boldsymbol{f}^*, \boldsymbol{\sigma}^*, \boldsymbol{\pi}^*)$, is a fixed price equilibrium of P2. The averaged iterate of step 4 remains feasible at every iteration.*

***Proof.***

a) *Seller step*. For each operator, the feasible set of step 3 is defined by Eqs. (42) – (43) and (45). It is nonempty by Assumption 1. It is compact and convex, because its constraints are linear in $\boldsymbol{Y}$. It does not depend on $\boldsymbol{L}$. The objective of step 3 is continuous in $(\boldsymbol{Y}, \boldsymbol{L})$, since $\sum_{a\in A_m} \delta_{ar}(\tau_a^0 L_a - c_a)$ is linear in $\boldsymbol{L}$. It is strictly concave in $\boldsymbol{Y}$, since the entropy terms are strictly concave and the rest is linear. So the maximizer $\tilde{\boldsymbol{Y}}(\boldsymbol{L})$ is unique for each $\boldsymbol{L} \in [0,1]^{|A_m|}$, and continuous in $\boldsymbol{L}$ by the maximum theorem. The feasible set contains a point with every $Y_r > 0$. At any $Y$ with some $Y_r = 0$, the derivative of $-\frac{1}{\alpha_{S,m}} Y_r(\ln Y_r - 1)$ is $+\infty$, so moving toward that point raises the objective. Hence $\tilde{Y}_r(\boldsymbol{L}) > 0$ for every $r$. $\tilde{\boldsymbol{Y}}$ is continuous on the compact set $[0,1]^{|A_m|}$, so $\tilde{Y}_r(\boldsymbol{L})$ is bounded below by a positive constant, uniformly in $\boldsymbol{L}$. Every owned link lies on some service (Assumption 2). So $y_a = \sum_r \delta_{ar} Y_r$ is bounded below by the same constant.
b) *Buyer step*. Fix $\boldsymbol{y}$ bounded below by this constant. Step 2 is the Lagrangian dual, at $\boldsymbol{y}$, of the strictly concave program that maximizes the buyer terms of Eq. (39) subject to Eqs. (40) – (41). That program is feasible since assigning $D_i$ to the outside links $a_{in}^0$ uses no owned link. Because $\boldsymbol{y} > \boldsymbol{0}$, this point is strictly feasible for Eq. (41). The feasible correspondence is therefore continuous in $\boldsymbol{y}$. The solution $\boldsymbol{f}$ is then unique, and by the maximum theorem it is continuous in $\boldsymbol{y}$, as are the link loads $f_a$.
c) *Composite map*. Restrict the feasible set of part (a) to $\boldsymbol{Y}$ bounded below by the constant. This set is still compact and convex. On it, $y_a = \sum_r \delta_{ar} Y_r$ is linear in $\boldsymbol{Y}$ and bounded away from zero. So $L_a = \min\left(1, \frac{f_a}{y_a}\right)$ is continuous, with values in $[0,1]$, by part (b). The composite of steps 2 and 3 is therefore a continuous map of this set into itself, by part (a). Brouwer's theorem gives a fixed point $\boldsymbol{Y}^*$. Since the set is convex, the averaged iterate of step 4 stays in it.

d) *Equilibrium interpretation*. At $\boldsymbol{Y}^*$, step 3 returns $\widetilde{\boldsymbol{Y}} = \boldsymbol{Y}^*$. Each operator's allocation solves its problem at $\boldsymbol{L}(\boldsymbol{Y}^*)$, with $\boldsymbol{\pi}^*$ the multipliers of Eq. (42). $\boldsymbol{f}^*$ solves the traveler problem at $y_a^* = \sum_r \delta_{ar} Y_r^*$, with $\boldsymbol{\sigma}^*$ the multipliers of Eq. (41). Complementary slackness gives $\sigma_a^*(y_a^* - f_a^*) = 0$. Travelers therefore face the effective price $\tau_a^0 + \sigma_a^*$, and the market clears at that price. ■

Theorem 3 establishes existence but not uniqueness. Algorithm 2' is a supply side day-to-day adjustment process in the sense of Cascetta and Cantarella (1991) and Cantarella and Watling (2016), with the operator's lagged load factor $\boldsymbol{L}^{(t)}$ playing the role assigned in those models to lagged perceived cost. A fixed point of the process map is by Theorem 3 a fixed price equilibrium, so the process converges to a fixed price equilibrium whenever it converges. Local stability of such processes holds when every eigenvalue $\lambda$ of the process Jacobian $J(\boldsymbol{Y}^*)$ of the map defined by steps 2-3 satisfies $\mathrm{Re}\,\lambda < 1$ asymptotically (Cantarella and Watling, 2016), where Re refers to the real part of the eigenvalue. Bie and Lo (2010) further characterize the attraction domain of a stable equilibrium in discrete-time day-to-day systems. Note that when the buyer and seller objectives are combined into a single coalitional decision-maker, this fixed price equilibrium becomes the lower-level equilibrium shown in Yang and Chow (2026), and hence serves as a generalization.

### 3.3. Inverse PURC-based model

Estimation of the parameters now assumes that segment-level link flows $f_{ia}^0$ are observed only on a subset of links, giving the set of observed cells $\mathcal{O}_F$, together with service capacities on $\mathcal{O}_S$, and aggregate link counts $f_a^0$ across all operated links are available to estimate the PURC-based model.

Additional notation to consider:

$z_{ia}$: vector of observable features of link $a \in A$ for segment $i \in Q$, with $z_{i,a_{in}^0} = 0$, so that $\varphi_{ia} = \boldsymbol{\theta}^T z_{ia}$ and $\varphi_{i,a_{in}^0} = 0$,

$\mathcal{O}_F \subseteq Q \times A$: set of sampled segment-link cells, obtained where fare-card, platform, or survey records identify the traveler segment traversing each link,

$\mathcal{O}_S \subseteq \{(m,r): r \in R_m, Y_r^0 > 0\}$: set of operated services with observed deployed capacities $Y_r^0$, where $y_a^0 = \sum_r \delta_{ar} Y_r^0$. For single-link services, this is the set of operated links with deployed capacities $y_a^0$,

$\mathcal{A}^+ = \{(i,a) \in \mathcal{O}_F : f_{ia}^0 > 0\}$ and $\mathcal{A}^- = \{(i,a) \in \mathcal{O}_F : f_{ia}^0 = 0\}$, where $\mathcal{A}_i^+$ is the set of links observed with positive flow for segment $i \in Q$ ,

$\mathcal{S} = \{(m,a): a \in A^{own}, f_a^0 \geq y_a^0(1 - \epsilon_b)\}$: binding set of links observed at capacity, where rationing prices may be positive, with tolerance $\epsilon_b \geq 0$,

$G$: set of supply-scale groups, with $g(m) \in G$ the group of operator $m$. Operators in a group share a scale, $\alpha_{S,m} \equiv \alpha_{S,g(m)}$,

$f_{ia}^0 \geq 0$: observed flow of segment $i \in Q$ on link $a \in A$ for $(i,a) \in \mathcal{O}_F$,

$p_a = \tau_a^0 + \sigma_a$: link level price facing travelers,

$\mu_{in} \equiv \alpha_B u_{in}$: reduced-form node potential, with $\mu_{i,s_i} = 0$,

$\varpi_{ia}^F > 0$: observation weight on $\mathcal{A}^+$, set to 1 in this study. If flows were observed with sampling error, e.g. Poisson counts, the inverse-variance choice $\frac{D_i^2}{\ell_a^2 f_{ia}^0}$ would apply,

$\rho_c, \rho_\pi \geq 0$: regularization weights,

$\vartheta_\sigma > 0$: ridge weight on the rationing prices,

$\kappa > 0$: bound on the rationing prices as a multiple of the largest observed fare.

**Assumption 3 (link flow observability)**. *Segment-level link flows $f_{ia}^0$ are observed without error on a subset of instrumented links: on each instrumented link, the flow $f_{ia}^0$ of every segment is recorded, and $\mathcal{O}_F \subseteq Q \times A$ is the resulting set of observed cells. The choice of instrumented links may be random and may* *depend on observable quantities such as tract volumes. Deployed service capacities $Y_r^0, r \in R$, maximum capacities $C_a^{max}$, and feature set $\boldsymbol{z_{ia}}$ are observed on all services and links $a \in A^{own}$ and are not sampled. Operating costs associated with each link $c_a$ for all links $a \in A^{own}$ are uncertain with a given prior $c^0$. The set $\mathcal{O}_S$ represents services with deployed capacity where the seller residual is defined.*

*Link fares $\tau_a^0$, totals $D_i$ and $F_m$, the node-link structure $t(a), h(a), b_{in}$ are known. Cells with $f_{ia}^0 = 0$ are therefore true zeros and are informative about the active support. Aggregate link flows $f_a^0$ are observed on all links $a \in A^{own}$ measured over the same time interval as $y_a^0$. These counts are a separate observation type from the sampled cells; they do not attribute flow to segments.*

Note that $c_a, \pi_a, \sigma_a, u_{in}$, and $\tilde{v}_m$ are now decision variables of the inverse problem IO1, where $\tilde{v}_m$ is the recentered value from Section 3.1. The socially optimal $\tau_a^*$ are determined by solving P2 after estimation. Rearranging the first order conditions in Eq. (47) on cells with positive flow, and substituting the effective link price $p_a$ for the clearing price $\tau_a^*$, we obtain Eq. (51),

$$\frac{\ell_a f_{ia}^0}{D_i} = \tilde{\theta}^T z_{ia} - \alpha_B p_a - \left(\mu_{i,t(a)} - \mu_{i,h(a)}\right), \qquad (i,a) \in \mathcal{A}^+ \tag{51}$$

The potentials enter in differences and are normalized by $\mu_{i,s_i} = 0$. Eq. (52) replaces $\tau_a$ with observed $\tau_a^0$, which may not be optimal. From observed data, we can define the link load factor $L_a = \min\left(1, \frac{f_a^0}{y_a^0}\right)$, with $y_a^0 = \sum_r \delta_{ar} Y_r^0$. A unit of capacity on service $r$ then earns the observed revenue $\sum_a \delta_{ar} \tau_a^0 L_a$, which replaces the clearing price in Eq. (52).

$$\frac{1}{\alpha_{S,m}} \ln \frac{Y_r^0}{F_m} = \sum_{a \in A_m} \delta_{ar}(\tau_a^0 L_a - c_a - \pi_a) - \tilde{v}_m, \qquad (m,r) \in \mathcal{O}_S \tag{52}$$

Eq. (51) enters the IO1 objective function as the buyer residual. For the idle alternative, Eq. (52) becomes $\frac{1}{\alpha_{S,m}} \ln \frac{y_{m0}^0}{F_m} = -c_{idle,m} - \tilde{v}_m$. The idle stock becomes $y_{m0}^0 = F_m - \sum_{r \in R_m} Y_r^0$. To preserve a quadratic form, the following reparameterizations are made: $\tilde{\boldsymbol{\theta}} \equiv \alpha_B \boldsymbol{\theta}$, $\tilde{c}_a \equiv \alpha_{S,g(m)} c_a$, $\tilde{\pi}_a \equiv \alpha_{S,g(m)} \pi_a$, $\tilde{\sigma}_a \equiv \alpha_B \sigma_a$, and $\beta_m \equiv \alpha_{S,g(m)} \tilde{v}_m$, with $\mu_{in} \equiv \alpha_B u_{in}$ as defined above. To preserve linearity, let $\tilde{c}_a = \alpha_{S,g(m)} c_a^0 + c_a^+ - c_a^-$. The inverse problem IO1 is expressed in Eqs. (53) – (60).

IO1:

$$\min_{\tilde{\theta}, \tilde{c}, c^+, c^-, \tilde{c}_{idle}, c_{idle}^+, c_{idle}^-, \tilde{\pi}, \alpha_B, \{\alpha_{S,g}\}_{g \in G}, \tilde{\sigma}, \mu, \beta} \mathcal{Z} = \sum_{(i,a) \in \mathcal{A}^+} \varpi_{ia}^F \left(\frac{\ell_a f_{ia}^0}{D_i} - \tilde{\boldsymbol{\theta}}^T \boldsymbol{z_{ia}} + \alpha_B p_a + \mu_{i,t(a)} - \mu_{i,h(a)}\right)^2 + \sum_{(m,r) \in \mathcal{O}_S} \varpi_r^S \left(\ln \frac{Y_r^0}{F_m} - \alpha_{S,g(m)} \sum_{a \in A_m} \delta_{ar} \tau_a^0 L_a + \sum_{a \in A_m} \delta_{ar}(\tilde{c}_a + \tilde{\pi}_a) + \beta_m\right)^2 + \sum_{m \in M} \varpi_m^0 \left(\ln \frac{y_{m0}^0}{F_m} + \tilde{c}_{idle,m} + \beta_m\right)^2 + \rho_c \left[\sum_{m \in M} \sum_{a \in A_m} (c_a^+ + c_a^-) + \sum_{m \in M} \left(c_{idle,m}^+ + c_{idle,m}^-\right)\right] + \rho_\pi \sum_{m \in M} \sum_{a \in A_m} \tilde{\pi}_a \tag{53}$$

s.t.

Reduced form definitions:

$$\tilde{c}_a = \alpha_{S,g(m)} c_a^0 + c_a^+ - c_a^-, \qquad \forall\, a \in A_m, m \in M \tag{54}$$

$$\tilde{c}_{idle,m} = \alpha_{S,g(m)} c_{idle,m}^0 + c_{idle,m}^+ - c_{idle,m}^-, \qquad \forall m \in M \tag{55}$$

Capacity slackness:

$$\tilde{\pi}_a = 0, \qquad \forall a \in A_m, m \in M: y_a^0 < C_a^{max} \tag{56}$$

Non-negativity

$$\tilde{\pi}_a \geq 0, \tilde{c}_a, c_a^+, c_a^- \geq 0, \qquad \forall\, a \in A_m, m \in M \tag{57}$$

$$\beta_m + \tilde{c}_{idle,m} \geq 0, c_{idle,m}^+, c_{idle,m}^- \geq 0, \qquad \forall m \in M \tag{58}$$

Normalizations:

$$\mu_{i,s_i} = 0, \qquad \forall i \in Q \tag{59}$$

$$\alpha_B, \alpha_{S,g} \geq \varepsilon, \qquad \forall g \in G, \qquad \varepsilon > 0 \tag{60}$$

Let the set of parameters, absent $\boldsymbol{\mu}$, be denoted collectively by $\boldsymbol{\Theta}$. Every residual is affine in $\boldsymbol{\Theta}$ after the reduced-form substitutions, so IO1 is a convex quadratic program; the buyer and seller blocks share no variable and separate. The seller block identifies costs only through the service sums $\sum_a \delta_{ar}\tilde{c}_a$, and only up to a common level per operator. Adding the same constant to every service sum of operator $m$ and to $\tilde{c}_{idle,m}$, and subtracting it from $\beta_m$, leaves every seller residual unchanged. This is because the operator's logit allocation depends only on cost differences among its alternatives. Reallocating a service's cost among its own links also leaves the residuals unchanged. Both the level and the split are pinned by the prior $\boldsymbol{c}^0$ through the $\rho_c$ term, and bounded below by $\tilde{c}_a \geq 0$. With single-link services, only the level is free. Estimates of $c_a$ and $\tilde{v}_m$, and the Remark 2 uses of $\tilde{v}_m$, therefore inherit the level of the prior.

The challenge with IO1 is the dimensionality of the node potentials, which can absorb nearly all of the residual when only a sample of segment-link flows is observed. Two remedies have been proposed. Fosgerau et al. (2026) project the potentials out through the cycle space of each segment's active subgraph. Rosenblad et al. (2026) retain the potentials and impose the observed inactive cells which supplies information that the active cells alone do not. We follow the second approach. Projection makes the potentials disappear, but an inactive cell can then be imposed only if both of its end nodes lie in the same component of the active forest. With the potentials kept, every observed inactive cell enters directly as a linear inequality on its reduced margin. The resulting buyer block is a sparse QP. Relative to Rosenblad et al. (2026), two features arise from the capacitated two-sided setting:

- The rationing price $\tilde{\sigma}_a$ is a decision variable only on the binding set $\mathcal{S}$, which is formed from observed counts and capacities. It carries the ridge $\vartheta_\sigma$ and is bounded by $\kappa\alpha_B\left(\max_a \tau_a^0\right)$, so that it cannot absorb residuals unrelated to capacity.
- The seller block shares no variables with the buyer block and is solved as a separate QP.

The full program, the buyer block of Eq. (53) together with the inequalities on $\mathcal{A}^-$, the bound on $\tilde{\sigma}_{\mathcal{S}}$, and $\tilde{\sigma}_a = 0$ off $\mathcal{S}$, is solved by Algorithm 3.

**Algorithm 3.** Solution to IO1 with buyer and seller QPs

***Input.*** Sampled flows $f_{ia}^0$ on $\mathcal{O}_F$; service capacities $Y_r^0$ on $\mathcal{O}_S$; observed fares $\tau_a^0$; aggregate link counts $f_a^0$; link features $z_{ia}$; priors $c_a^0, c_{idle,m}^0$; totals $D_i, F_m$; caps $C_a^{max}$; node-link structure $t(a), h(a), b_{in}$; length weights $\ell_a$; services $\{R_m\}$ with incidence $\delta_{ar}$; objective weights $\rho_c, \rho_\pi$; rationing-price weight $\vartheta_\sigma > 0$, premium bound $\kappa$; binding tolerance $\epsilon_b$; floor $\varepsilon > 0$; clearing iteration budget $N_{clear}$; map $g(m)$ assigning operator $m$ to supply-scale group $g \in G$.

***Output***. $\hat{\theta}, \hat{c}, \hat{c}_{idle}, \hat{\pi}, \hat{\alpha}_B, \{\hat{\alpha}_{S,g}\}_{g\in G}, \hat{u}, \hat{v}, \hat{f}, \hat{Y}, \hat{\sigma}; \tau^*, \Delta$.

1. *Preprocess.* Partition $\mathcal{O}_F$ into $\mathcal{A}^+$ and $\mathcal{A}^-$ by the sign of $f_{ia}^0$. Form the link capacities $y_a^0 = \sum_r \delta_{ar} Y_r^0$, and fix $\tilde{\pi}_a = 0$ on links with $y_a^0 < C_a^{max}$. Form $l_r^S = \ln\left(\frac{Y_r^0}{F_m}\right)$, the load factors $L_a = \min\left(1, \frac{f_a^0}{y_a^0}\right)$, and the idle stock $y_{m0}^0 = F_m - \sum_{r\in R_m} Y_r^0$, and $q_{ia} = \frac{\ell_a f_{ia}^0}{D_i}$ on $\mathcal{A}^+$; set $\varpi_r^S = Y_r^0, \varpi_m^0 = y_{m0}^0, \varpi_{ia}^F = 1$.
2. *Seller block.* Solve the QP:

$$\min_{\tilde{c},c^+,c^-,\tilde{c}_{idle},c^+_{idle},c^-_{idle},\tilde{\pi},\{\alpha_{S,g}\},\beta} \mathcal{Z}^S$$

$$= \sum_{(m,r)\in O_S} \varpi_r^S \left( l_r^S - \alpha_{S,g(m)} \sum_a \delta_{ar}\tau_a^0 L_a + \sum_a \delta_{ar}(\tilde{c}_a + \tilde{\pi}_a) + \beta_m \right)^2$$

$$+ \sum_{m\in M} \varpi_m^0 \left( \ln\frac{y_{m0}^0}{F_m} + \tilde{c}_{idle,m} + \beta_m \right)^2 + \rho_c \left[ \sum_{m\in M}\sum_{a\in A_m} (c_a^+ + c_a^-) + \sum_{m\in M} \left( c^+_{idle,m} + c^-_{idle,m} \right) \right]$$

$$+ \rho_\pi \sum_{m\in M}\sum_{a\in A_m} \tilde{\pi}_a$$

s.t.

Eqs. (54) – (58), and

$$\alpha_{S,g} \geq \varepsilon, \qquad \forall g \in G$$

Descale for each $a \in A_m$: $\hat{c}_a = \frac{\hat{\tilde{c}}_a}{\hat{\alpha}_{S,g(m)}}$, $\hat{c}_{idle,m} = \frac{\hat{\tilde{c}}_{idle,m}}{\hat{\alpha}_{S,g(m)}}$, $\hat{\pi}_a = \frac{\hat{\tilde{\pi}}_a}{\hat{\alpha}_{S,g(m)}}$, $\hat{v}_m = \frac{\hat{\beta}_m - \ln F_m}{\hat{\alpha}_{S,g(m)}}$.

3. *Support of the rationing prices*. Form the binding set from the observed aggregate counts and deployed capacities, $\mathcal{S}$, and set $\tilde{\sigma}_a = 0\ \forall (m,a) \notin \mathcal{S}$.
4. *Buyer block.* Solve the QP:

$$\min_{\tilde{\theta},\alpha_B\geq\varepsilon,\tilde{\sigma}_{\mathcal{S}},\mu} \sum_{(i,a)\in\mathcal{A}^+} \varpi_{ia}^F \left( q_{ia} - \tilde{\theta}^T z_{ia} + \alpha_B \tau_a^0 + \tilde{\sigma}_a + \mu_{i,t(a)} - \mu_{i,h(a)} \right)^2 + \vartheta_\sigma \|\tilde{\sigma}_{\mathcal{S}}\|_2^2$$

s.t.

$$\tilde{\theta}^T z_{ia} - \alpha_B \tau_a^0 - \tilde{\sigma}_a - \left( \mu_{i,t(a)} - \mu_{i,h(a)} \right) \leq 0, \qquad \forall (i,a) \in \mathcal{A}^-$$

$$0 \leq \tilde{\sigma}_a \leq \kappa \left( \max_a \tau_a^0 \right) \alpha_B, \qquad \forall (m,a) \in \mathcal{S}$$

$$\mu_{i,s_i} = 0, \qquad \forall i \in Q$$

*Descale* $\hat{\theta} = \frac{\hat{\tilde{\theta}}}{\hat{\alpha}_B}$, $\hat{\sigma}_a = \frac{\hat{\tilde{\sigma}}_a}{\hat{\alpha}_B}$, $\hat{\varphi}_{ia} = \hat{\theta}^T z_{ia}$.

5. *Fixed-price allocation*: Solve P2($\boldsymbol{\tau^0}$) using Algorithm 2' at $\left(\hat{\boldsymbol{\varphi}}, \hat{\boldsymbol{c}}, \hat{\boldsymbol{c}}_{\boldsymbol{idle}}, \hat{\alpha}_B, \{\hat{\alpha}_{S,g(m)}\}_{m\in M}, \boldsymbol{\tau^0}\right)$. This returns $\hat{\boldsymbol{u}}, \hat{\boldsymbol{Y}}$, and every $\hat{f}_{ia}$ including unsampled cells, and the rationing prices $\boldsymbol{\sigma}^{(P2)}$ implied by those primitives.
6. *Clearing prices and wedge.* Solve P2 using Algorithm 2 at the same primitives, to a residual tolerance rather than an iteration cap, to obtain $\tau_a^*, f_{ia}^*$, and the wedge $\Delta_a = \tau_a^0 - \tau_a^*$. Report the clearing solve's terminal residual and whether it met tolerance within $N_{clear}$.

---

**Remark 5. Identification.** *With* $\boldsymbol{f}$ *fixed at* $\boldsymbol{f^0}$*, the buyer block of IO1 is convex but not strictly convex in* $(\boldsymbol{\Theta}, \boldsymbol{\mu})$*. Adding a constant to the potentials of every node in a connected component of a segment's active graph on* $\mathcal{A}^+$ *leaves every difference* $\mu_{i,t(a)} - \mu_{i,h(a)}$ *unchanged, for any component other than the one containing the sink* $s_i$*. Such potentials are not unique, but they do not affect* $\boldsymbol{\Theta}$*. Beyond this,* $\boldsymbol{\Theta}$ *is identified only through combinations of residuals in which the potentials cancel, i.e., through cycles of each segment's active graph. Identification therefore requires each segment to offer alternative routes with distinct link features. A segment whose active cells form a tree, with no alternative routes, contributes nothing to* $\boldsymbol{\Theta}$ *through Eq. (51). Some directions of* $\tilde{\boldsymbol{\theta}}$ *may be confounded with the rationing prices, if their effect on the cycles can be reproduced by adjusting* $\tilde{\sigma}_a$ *on the binding links in* $\mathcal{S}$*. Directions that remain confounded even when every cell is observed are structural. They are resolved by the bound and ridge on* $\tilde{\boldsymbol{\sigma}}$*, or fixed by normalization. A sampled design is admissible if it loses no identification beyond those structural directions. This depends only on* $O_F$ *and* $\mathcal{S}$*, so it can be checked before estimation. The inequalities on* $\mathcal{A}^-$ *add information beyond the cycles. They can bound directions that Eq. (51) leaves free, which is why observed zero cells improve the estimate even when the cycles alone already identify* $\boldsymbol{\Theta}$*. On the seller side, the scale of group* g *is the slope of* $\ln\left(\frac{y_a^0}{F_m}\right)$ *on the observed service revenue* $\sum_{a\in A_m} \delta_{ar}\tau_a^0 L_a$ *across the services of its* *operators, with $\beta_m$ absorbing each operator's level. It is identified when the group has at least two services in $O_S$ below their caps with different margins.*

**Remark 6. Optimality.** *Algorithm 3 returns a global minimizer of IO1. Every residual of IO1 is affine in the decision variables after the reduced-form substitutions, and every constraint is linear. IO1 is therefore a convex QP, and any point satisfying its KKT conditions is a global minimizer. The buyer and seller blocks share no decision variables. The objective is the sum of the two blocks, and each constraint involves only one of them. Minimizing IO1 is therefore equivalent to minimizing each block separately, and the union of the two block minimizers is a minimizer of IO1. Steps 2 and 4 solve exactly these two blocks to optimality, each as a QP. Steps 5 and 6 are not part of the estimation. They evaluate the estimated primitives with the forward models. Step 5 invokes Algorithm 2'. Each of its blocks is strictly convex, and its fixed point exists by Theorem 3, but the composite is not itself a convex program.*

**Remark 7. Other pricing instruments**. *The estimated model can also evaluate pricing instruments other than per-link cost transfers. Under a flat subscription, for example, subscribed travelers face no marginal fare on owned links, so the fixed-price equilibrium of Algorithm 2' is re-solved with link prices $\sigma_a$ in place of $\tau_a^0 + \sigma_a$ for that traveler group. Because the equilibrium of Eqs. (39) – (45) is already the entropy-regularized system optimum (Remark 1), such instruments are second best: they can raise welfare only through margins outside the price system, such as platform participation or recovery of operator fixed costs. Fares set per service or per origin-destination pair are covered in the same way when they are additive over links, as a boarding charge or a per-mile tariff is. When they are not, they restrict the link price system and are likewise second best. Their design is left to future work.*

## 4. Numerical tests

The goal of the first section is to verify the models P1 and P2 to illustrate how they can be used. Using a toy network, we first solve P1 and P2 to compare the flows, confirming the output dual variables correspond to the primal accordingly and examine each operator's marginal contribution to the market. We show how as the parameters $\alpha_B$ and $\alpha_{S,m}$ increase toward infinity, the solutions converge toward a deterministic result. Finally, we show how when one operator's $F_m$ increases, it impacts the other operator's outcome.

The second set of tests deals with IO1 and applies to a much larger network, the LIRR morning commute network. The purpose is to benchmark the effectiveness of Algorithm 3 in solving IO1 assuming a modeler is provided with observed flows in which an oracle is aware of all flows but the modeler may only be sampling a portion. The underlying model and parameters used to generate the flows are intentionally designed under a different specification and kept hidden to represent specification noise. Metrics include computation time, average errors compared with the flows and the rationing prices derived from the observed data as the sampling rate increases. Then P2 is used to recover the pricing wedge and gross surpluses under these sampling rates to compare them against the oracle. An ablation study is conducted to evaluate the contribution of the three components of the IO1 algorithm: restricting the use of inactive cells restricted to only reachable cells from a cycle projection to compare to Fosgerau et al. (2026), restricting all inactive cell use to confirm the conclusion from Rosenblad et al. (2026), and restricting the rationing prices to zero to test the importance of modeling the fixed price equilibrium with rationing prices. Finally, the optimal tolls are mapped back to the LIRR zonal fare structure to compare against the synthesized observed toll schedule.

All results were obtained on a Dell XPS 8930 desktop with an Intel Core i9-9900K CPU (3.6 GHz, 8 cores, 16 threads) and 64 GB RAM, running Microsoft Windows Home, using Python 3.1.3 with NumPy 2.5.2, SciPy 1.18.0, and pandas 3.0.5. The dual problems of Algorithms 1, 2, and 2' are solved with SciPy's L-BFGS-B implementation (Byrd et al., 1995), and the seller and buyer QPs of Algorithm 3 are solved with CVXPY 1.9.3 (Diamond and Boyd, 2016) using the Clarabel interior-point solver 0.11.1 (Goulart and Chen, 2026). The codes and input data for the two test networks are provided in Github (Chow, 2026).

### 4.1. Toy example

The input parameters for the toy network are provided in Fig. 1. Note that the feature values are relative to the "other" options, i.e. IVT of link 1 is 20 minutes longer than that of link 5 (the other option for connecting node 1 to 2). The design of this toy network is meant to represent a residential suburb (node 1) connecting to a transit hub (node 2) and continuing to a downtown (node 3). There are two transport providers here: RED (dot-dash) and BLUE (solid). Other options are represented by GRAY (dash). RED is meant to have a much smaller capacity operation ($F_R < F_B$) as well as max capacities at links 3 and 4, but perform more efficiently than BLUE (think some on-demand service compared to BLUE's commuter bus). The $\boldsymbol{\theta}$ are normalized to convert the features to units of \$ in the utility functions, and set up to reflect a $VOT_{IVT} =$ \$$15/hr$ and $VOT_{OVT} =$ \$$30/hr$.

| $i$ | $(o_i, s_i)$ | $D_i$ (pax/hr) | $K_i$ |
|---|---|---|---|
| 1 | (1,2) | 400 | $\{l_1\}, \{l_3\}$; other: $\{l_5\}$ |
| 2 | (2,3) | 600 | $\{l_2\}$; other: $\{l_6\}$ |
| 3 | (1,3) | 800 | $\{l_4\}, \{(l_1, l_2)\}, \{(l_3, l_2)\}$; other: $\{l_7\}$ |

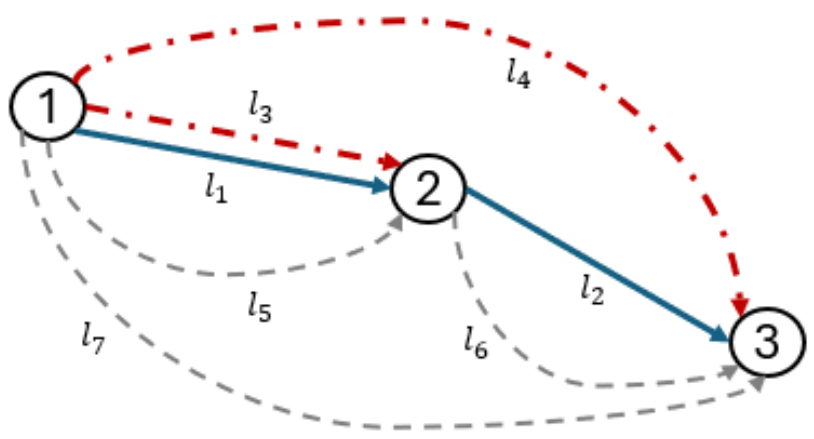

$F_B = 1200, F_R = 250$

For P1/P2: $\theta = (\theta_d, \theta_{IVT}, \theta_{OVT}) = (1.20, -0.25, -0.50)$,

$\alpha_B = 1.0, \alpha_S = 0.8, c_{idle,R} = c_{idle,B} = 0$

| link | $\ell_a$ (mi) | $IVT$ (min) | $OVT$ (min) | $\varphi_a$ | $C_{ma}^{max}$ (pax/hr) | $c_{ma}$ (\$/unit) | $\tau_{ma}^0$ (\$) |
|---|---|---|---|---|---|---|---|
| $l_1$ | 8 | 20 | 5 | 2.10 | 900 | 1.40 | 2.75 |
| $l_2$ | 10 | 25 | 5 | 3.25 | 900 | 1.70 | 2.75 |
| $l_3$ | 8 | 12 | 2 | 5.60 | 200 | 0.80 | 3.50 |
| $l_4$ | 18 | 30 | 2 | 13.10 | 150 | 1.50 | 6.00 |
| $l_5$ | 8 | 0 | 0 | 0 | - | - | - |
| $l_6$ | 10 | 0 | 0 | | | | |
| $l_7$ | 18 | 0 | 0 | | | | |

Fig. 1. Inputs for toy network.

#### 4.1.1. P1 equilibrium on the toy network

First, we verify that P1 is formulated correctly and that Algorithm 1 converges to the solution. It terminates on iteration $t = 19$ using $\epsilon = 1 \times 10^{-7}$, with $W^* = -13391.852$, and a gross surplus of $G^{P1} =$ \$3454.16. The trajectories of the natural residual $\varrho$ and optimality gap are shown in Fig. 2. Wall time is 38.5 ms.

The output flows, capacities, and prices are shown in Fig. 3. We see that the BLUE operator captures more of the flow, but RED operator is the one that is binding with link $l_4$. Summing up the capacities allocated, RED allocates 99.8% of its capacity to this market while BLUE allocates 85.0%, leaving the rest idle.

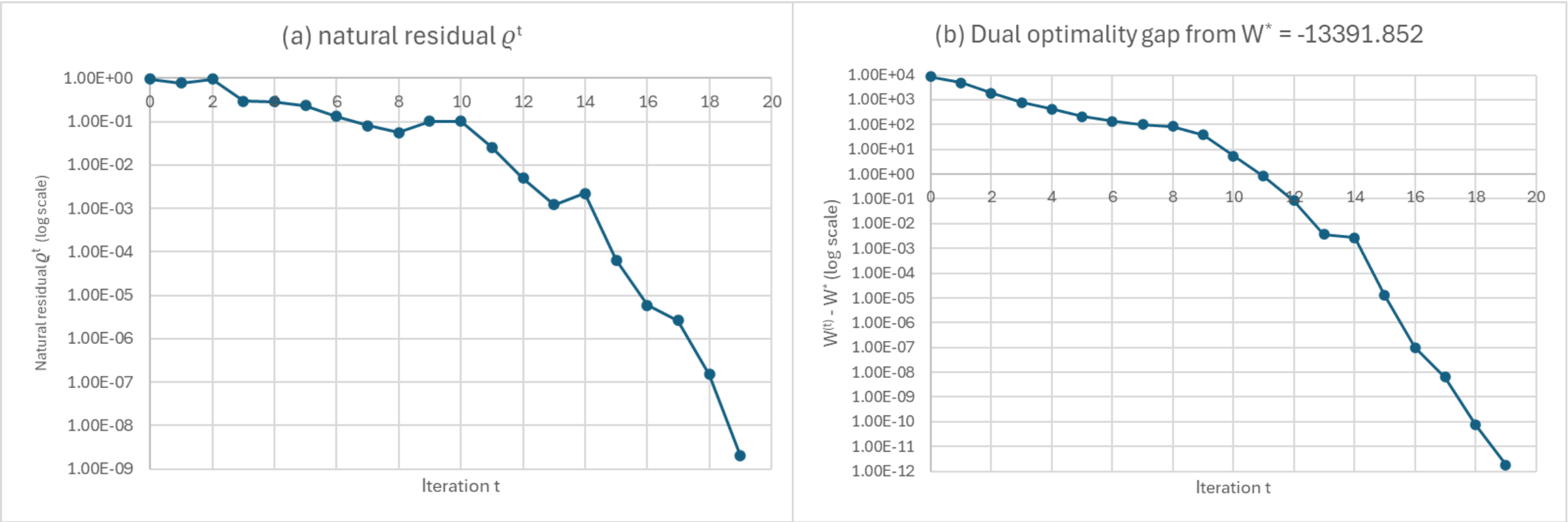

Fig. 2. Trajectory of Algorithm 1 applied to P1 for toy network: (a) natural residual, and (b) dual objective optimality gap.

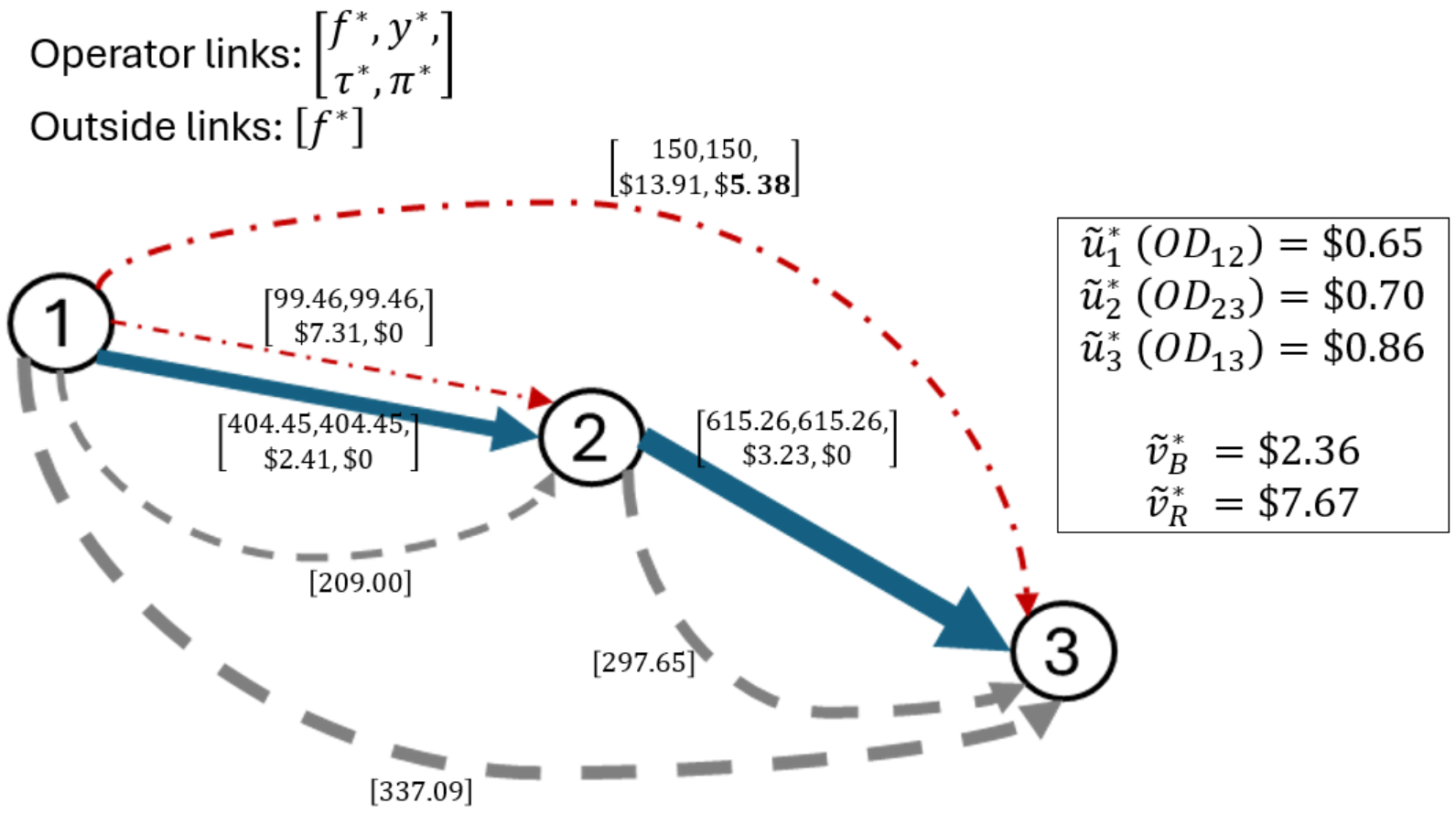

Fig. 3. Flows, capacities, and prices from solution to P1 (link widths scaled to flows).

The dual variables suggest that OD (1,3) travelers gain the most from this market with a per-traveler logsum of $\tilde{u}_3^* = \$0.86$. On the supply side, the marginal social surplus of capacity is $\tilde{v}_R^* = \$7.665$ for RED and $\tilde{v}_B^* = \$2.369$ for BLUE, so an additional unit of RED capacity is worth \$5.30 more than an additional unit of BLUE capacity. Capacity expansion targeted at RED dominates in this example. These values also determine each operator's deployment. The reserve fraction (with $c_{idle,m} = 0$) is $\frac{y_{m0}^*}{F_m} = \exp\left(-\alpha_{S,m}\left(c_{idle,m} + \tilde{v}_m^*\right)\right) = \exp(-0.8 \times 7.665) = 0.2\%$ idle for RED, and 15.0% idle for BLUE.

### *4.1.2. P2 equilibrium on the toy network*

Now consider using the PURC-based framework to assign the flows, capacities, and prices. The output changes slightly, noting that PURC is not restricted to Independence of Irrelevant Alternatives conditions as MNL. The results are obtained in 21 iterations (wall time 6.2 ms), with a dual objective of $W^* = -\$10{,}038.82$ (cannot be compared with P1's), and a gross surplus (can be compared with P1's result) of $G^{P2} = \$3472.44\ (+0.53\%\ vs\ P1)$. The 0.53% difference represents differences between two behavioral models, not an error. The output flows, capacities, and prices are plotted in Fig. 4. Algorithm 2 is six times faster in wall time, even for such a small network where the number of paths is limited.

P2 prices everything higher, and the gap widens with scarcity. This is likely because the quadratic perturbation is less dispersive than the Shannon entropy for the same parameters, so more travelers want to use the market (and fewer on outside options), which is reflected in higher pricing. Both operators end up allocating more capacity as well, with RED allocating up from 99.8% to 99.9% of its capacity to this market while BLUE shifting from 85.0% to 87.1% in P2, leaving the rest idle.

At the observed fares, Algorithm 2' is further used to solve $P2(\boldsymbol{\tau}^{\mathbf{0}})$ in 12 iterations to tolerance (wall time 37.2 ms), with a $G^{P2}(\boldsymbol{\tau}^{\mathbf{0}}) = \$3141.93$, which is 9.5% less than the gross surplus of the social optimum with a gap of $330.51.

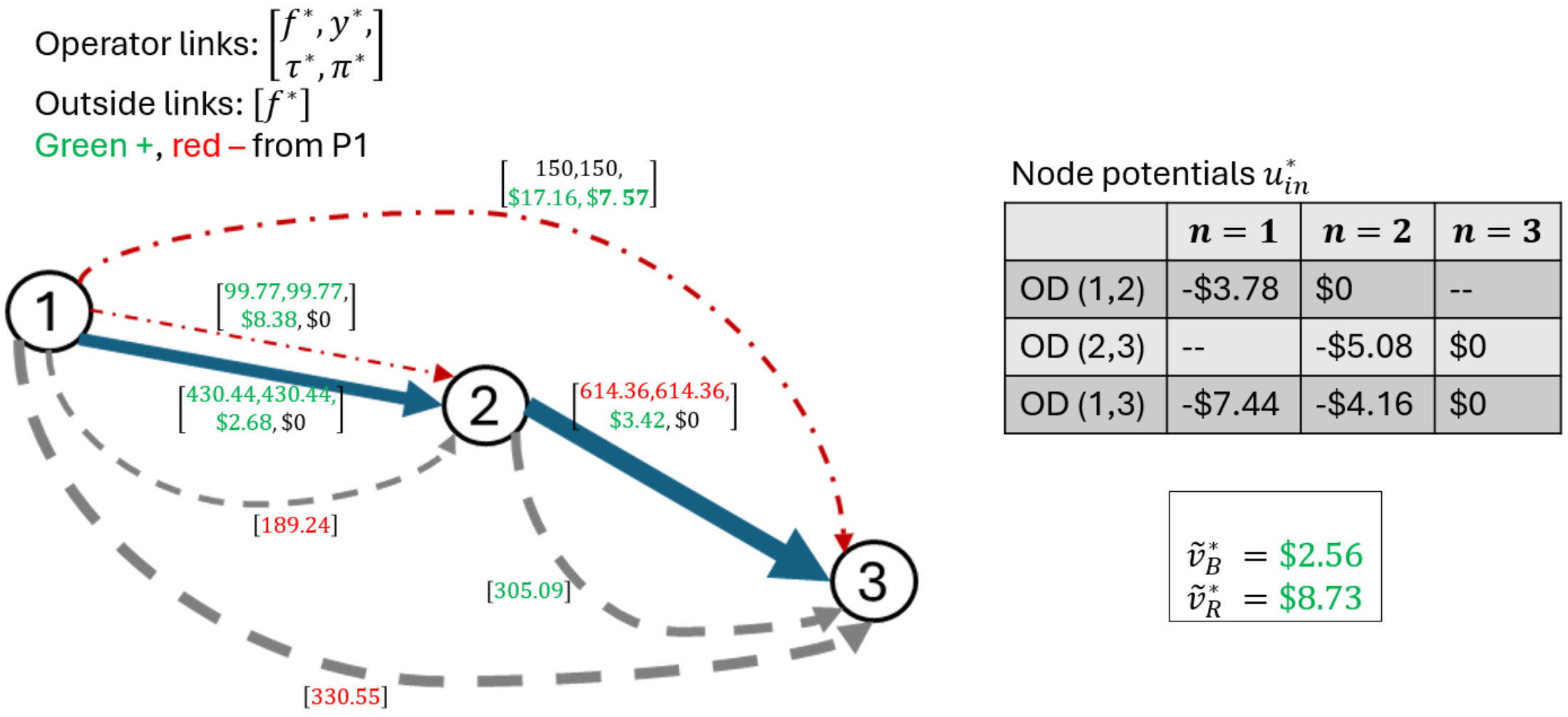

| | $n = 1$ | $n = 2$ | $n = 3$ |
|---|---|---|---|
| OD (1,2) | -$3.78 | $0 | -- |
| OD (2,3) | -- | -$5.08 | $0 |
| OD (1,3) | -$7.44 | -$4.16 | $0 |

Fig. 4. Flows, capacities, and prices from P2 with changes from P1 noted.

**Remark 8. The regularized core is a singleton.** *At* $\alpha_B = 1, \alpha_{S,m} = 0.8$ *the dual of P1 has a unique minimizer, so the regularized* $\mathcal{C}$ *of Corollary 1 is a single point. This follows from Theorem 1(2). Every owned link lies on some path containing no other owned link. For example,* $l_1$ *and* $l_3$ *for OD(1,2),* $l_2$ *for OD(2,3), and* $l_4$ *for OD(1,3). By contrast, the core of the unregularized game need not be a single point; the entropy terms select one element from it.*

**Remark 9. Deterministic limits.** *As* $\alpha_B, \alpha_{S,m} \to \infty$, *the entropy and quadratic perturbations vanish, and P1 and P2 both reduce to the linear program that maximizes* $\sum_{i \in Q} \sum_{k \in K} \Phi_{ik} x_{ik} - \sum_{m \in M} \sum_{a \in A_m} c_a y_a$ *subject to* $Eqs. (21) - (25)$, *or equivalent link form for P2. Solving this LP directly for the toy network gives* $G = \$3{,}825$. *Both operators are fully deployed, with* $l_2$ *and* $l_4$ *at their caps, and 300 OD (1,2) and 350 OD (1,3) travelers remain on the outside option. For* $\alpha_B, \alpha_{S,m} \geq 500$, *both regularized models reach this same value,* $G^{P1} = G^{P2} = \$3825$, *confirming they recover the deterministic assignment game in the limit (Corollary 1(iii)).*

**Remark 10. Cross-operator externality**. *The effect that each operator's endowment has on one another is demonstrated by varying* $F_R \in \{50, 55, \ldots, 345\}$ *and solving P2 each time. The results of the sweep are plotted in several graphs in Fig. 5.*

*In Fig. 5(a), both operators exhibit monotonically decreasing curves as RED's endowment increases. Furthermore, there is a big shift in RED at 155 due to the presence of that binding capacity at link* $l_4$. *BLUE is not operating at capacity anywhere and as a result the value of adding to its endowment is quite low by comparison. Fig. 5(b) shows that the change in* $F_R$ *affects all links' allocations among both operators. Fig. 5(c) shows the externality of RED's endowment on BLUE's capacity allocation. Capacity at* $l_4$ *is prioritized until* $F_R = 155$ *at which point the* $l_4$ *capacity is binding, after which additional capacity is loaded onto* $l_3$ *which competes with* $l_1$ *but complements* $l_2$, *thus resulting in the divergence in their marginal effects.*

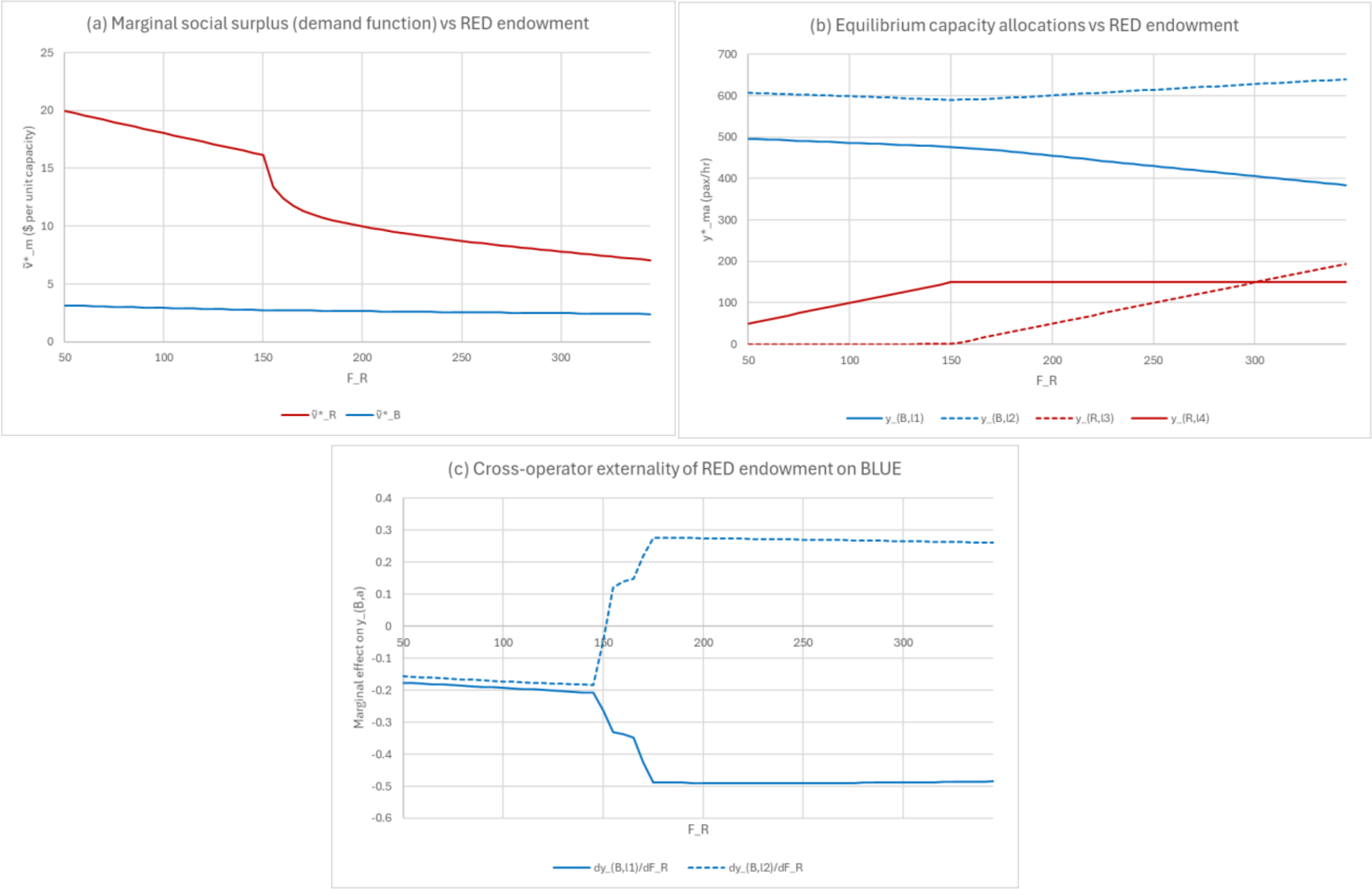

Fig. 5. Illustration of the effects of $F_R \in [50,345]$, via (a) marginal social surplus of both operators, (b) both operators' equilibrium capacity allocations, and (c) cross-operator externality on BLUE.

## *4.2.LIRR Example*

### *4.2.1. Data and experimental design*

Whereas Section 4.1's toy network is used to illustrate the forward models P1 and P2, the goal of this section is to verify that the proposed Algorithm 3 works as intended and to evaluate its effectiveness in inferring parameters around an aggregate network assignment game from observed data for subsequent analysis and design. For that purpose, we use the LIRR network. A set of synthetic flows is generated by a model based deliberately on a different specification to mimic the real data where true parameters are unknown. The modeler observes fares, aggregate link counts, deployed service capacities, together with a sample of disaggregate segment-level flows, and nothing else. The data-generating model is the fixed-price equilibrium of P2 with a richer utility specification (97 station constants) than the estimated model. The oracle knows all the true $\theta$ parameters, but it would not make sense to compare those parameters with the misspecified model. Instead, despite not knowing the model specification in advance, we want to see whether aggregate metrics like gross surplus and optimal tolls can still approach the true values. One should expect:

- The inverse problem should be able to estimate parameters to fit the model to the flows, and should generally improve as the sampling rate increases.
- The inverse problem is able to recover the general rationing price patterns from the observed data under observed tolls $\boldsymbol{\tau}^{\mathbf{0}}$. Capturing the patterns shows that the inverse problem can account for capacity effects on travelers' route choices under non-optimal tolls.

- Whether the inverse problem can get representative values of gross surplus or pricing wedge despite having a different specification of the model, and whether those estimates show stabilization with the sampling rate.

The LIRR instance represents the AM morning peak period commuter market from Long Island to New York City (NYC), served by three operators: LIRR commuter rail network, the station parking lots, and a local feeder transit to those stations. Travelers can choose to take the local transit or drive themselves to a nearby station to take LIRR, or to avoid the system entirely and drive themselves to NYC (similar construct to Yang and Chow, 2026, but applied across the whole network instead of just the vicinity of Ronkonkoma Station). The network and demand are constructed from public data where available, and visualized in Fig. 6.

- *Size*. The network has 897 nodes, consisting of 671 census tracts, 123 LIRR stations, 97 parking nodes, and 6 NYC destination zones, with 8,320 directed links of which 2,275 are operator-owned. Each tract is connected to three candidate stations by access links from each tract to each of its three candidate stations (4,026 directed links); 326 drive-access and 124 transit-access links of the 2,013 tract-to-station links carry flow. Demand forms 3,990 OD pairs (665 tracts × 6 destination zones; six tracts have no commuters to these zones), grouped into 6 destination-based segments. Because each segment can use only links on routes to its own destination, 29,455 of the 49,920 segment-link cells are available, and 4,673 of these carry positive flow, a sparsity of 0.84.
- *Demand*. Trip productions come from LEHD LODES8 (the ny_od_main_JT00_2022 file) (U.S. Census Bureau, 2025), which includes 410,320 Long Island residents working in NYC by home tract. Given the small portions of Bronx and Staten Island destinations, they are dropped and the remainder scaled to 53.5% Manhattan, 28.9% Queens, and 17.6% Brooklyn. Tract centroids and GEOIDs are from the 2025 Census Gazetteer.
- *Supply and calibration targets*. There are three operators: LIRR provides the rail service ($F_{LIRR} = 179{,}520$ seats per AM peak), LOCAL operates the feeder transit lines from census tracts to stations ($F_{LOCAL} = 88{,}948$), and PARK operates the station parking lots ($F_{PARK} = 62{,}327$). LIRR deploys seats on 481 services, each a train pattern from an origin station to an NYC terminal (up to 34 links). LOCAL and PARK allocate capacity link by link, so each of their 2,050 feeder links and 97 lots is a single-link service. The instance is therefore a mixed line-link system under the single formulation P2. The station boarding profile is taken from the MTA 2012-2014 LIRR Origin and Destination Report (westbound, AM peak, boardings) (MTA, 2022), rescaled to the AM peak control total implied by the MTA 2025 Annual Ridership Report (MTA, 2025). Operating expense is $27.90 per vehicle revenue mile or $0.254 per seat-mile at 110 seats per car, so $c_a$ = $0.254 per seat-mile times the length of link $a$ (FTA, 2023). At the observed fares, AM peak fare revenue covers 94% of the cost of the seats running at 45% average occupancy. Systemwide, LIRR fares recover 27% of operating expense.

  Station names, branch topology and fare zones are real; a zone-gap fare schedule is used based on MTA (2026) (translated to link prices as described below). Station parking charges are assumed, decaying with distance to Manhattan over $0.78 - $6.26. The seller scales are set to $\alpha_{S,LIRR} = 0.25$, and $\alpha_{S,LOCAL} = \alpha_{S,PARK} = 4$, with LOCAL and PARK in one estimation group, since 94 of the 97 lots are at their cap.

To synthesize the flows, we use the published fares translated to the link level as a two-part tariff, i.e. a charge of $8.64 on the links entering an NYC terminal plus $0.233 per mile. The ground truth is the fixed-price equilibrium of P2 at these fares, solved by Algorithm 2'. The utility function includes five time coefficients at -$0.367/min on the rail, drive-access, transit-access, egress, and outside legs; a non-fare cost coefficient of -1.0; 97 station constants -$16.23 to +$5.35 placed on boarding links to fit to observed boardings; and a single drive constant of +$5.54 on the outside links. The synthesized flows produce 71,930 AM peak boardings against an observed 72,130, with a correlation of 0.967 and a rail mode share of 28.3% against an observed 28.4%.

To avoid allowing the model to reproduce the utility function exactly, we assume a different specification model for the inverse problem than the synthesized data, under the assumption of having no prior specification knowledge. In the inverse problem, we leave out the 97 station constants used in the synthetic data generation, which means that our estimated model will have noise. As such, we do not focus on the recovery of the $\theta$ but rather on the ability of the inverse model to fit to flows and rationing prices.

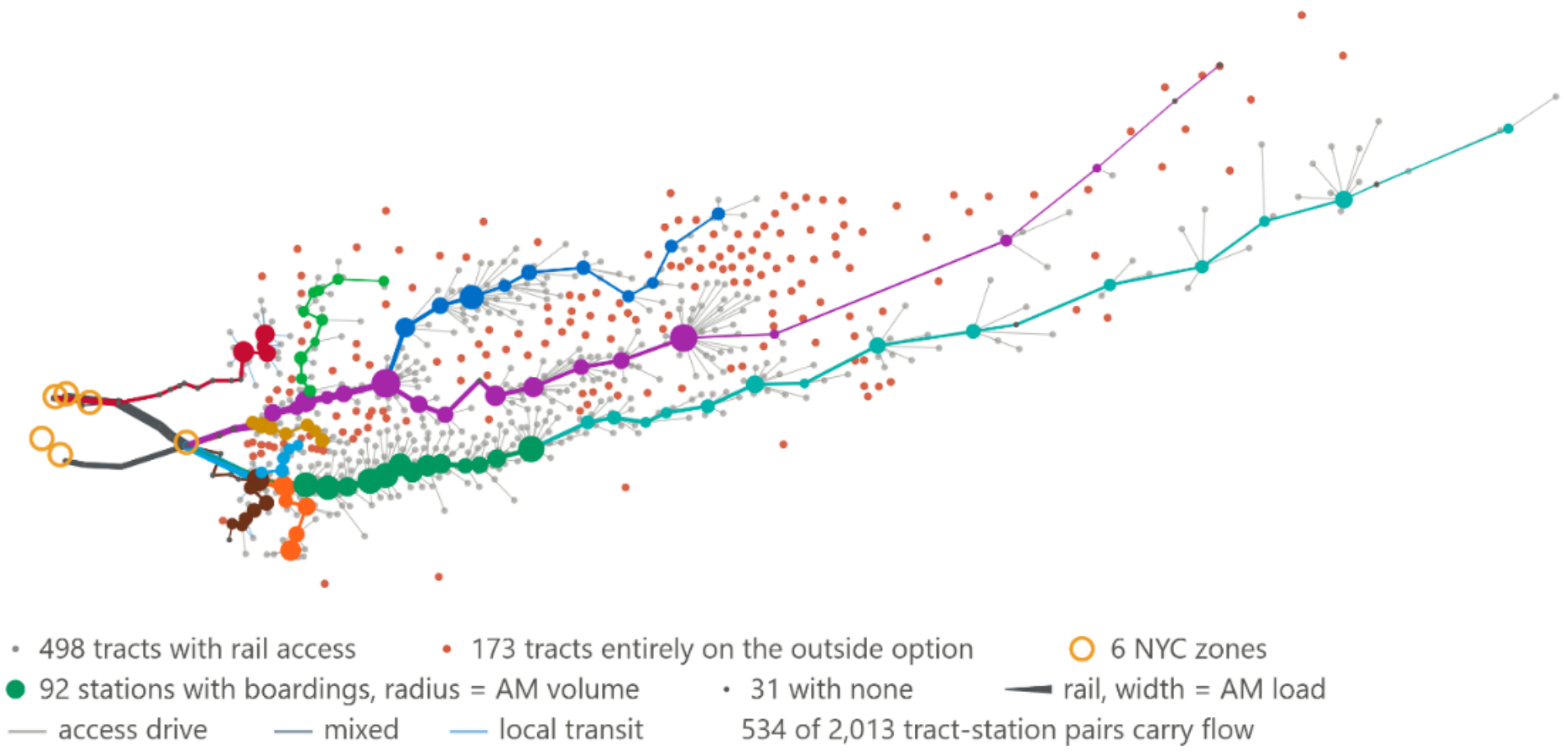

Fig. 6. The LIRR network with census tracts and synthetic links and flows constructed from public data and an assumed model. The synthetic data is available on Github (Chow, 2026).

For the inverse problem, we assume that all deployed service capacities $Y_r^0$ and aggregate flows $f_a^0$ are *observed* to the modeler. The oracle further knows the *actual* $f_{ia}^0, \sigma_a^0, \boldsymbol{\theta}$, and the derived aggregate metrics like $G(\boldsymbol{\tau}^{\mathbf{0}}), G(\boldsymbol{\tau}^*)$, and $\Delta$. We fix $\alpha_B = 1$ as an identifying restriction on the money metric. Under the misspecified model, fare variation within cycles is largely the per-mile tariff, which is collinear with rail time, and parking-fee variation is confounded with the omitted station constants, so a free $\alpha_B$ is biased downward. We set $\rho_c = 100$ to indicate a high analyst's confidence in the cost prior, and set $\rho_\pi = 1$. The rationing price weight is set to $\vartheta_\sigma = 10^{-3}$ is selected on a held-out validation link set and held fixed across sampling rates. The synthetic ground truth metrics are reported in Table 1.

Table 1. Summary of synthesized ground truth values

| Metric | Value |
|---|---|
| Actual gross surplus, $\boldsymbol{G}(\boldsymbol{\tau}^{\mathbf{0}})$ | \$894,503 |
| Social optimal gross surplus, $\boldsymbol{G}(\boldsymbol{\tau}^*)$ | \$1,965,043 |
| Deadweight loss (DWL) | \$1,070,540 |
| DWL as percent of social optimal gross surplus | 54.5% |
| Mean \| max optimal toll $\boldsymbol{\tau}^*$ | \$0.31 \| \$6.56 |
| Mean absolute pricing wedge $\overline{\lvert\Delta_{\boldsymbol{a}}\rvert}$ | \$2.45 |

Active cells inform the taste parameters only through cycles of a segment's observed active subgraph, and observed zero cells add inequalities, so links are sampled in bundles. Five sampling rates are assumed: 10%, 20%, 30%, 40%, and 50%, each with five replications (seeds). The sampling rate is the fraction of the 8,320 links that are instrumented. A link observation contributes to the inverse problem only by closing a cycle in a segment's observed active subgraph, so links are sampled in bundles by origin census tract, which keeps the access alternatives from a tract observed together. Tracts are drawn without replacement with probability proportional to the square root of tract volume, a compromise between simple random sampling and volume-proportional sampling. Bundles are added until the target share of links is reached. Rail and egress links are observed at every rate. On each instrumented link, the flows of all six segments are recorded exactly. The only source of randomness is which links are instrumented. Draws are nested. For each seed, the sample at a lower rate is a subset of the sample at a higher rate. The LIRR network and synthesized data are available on Github (Chow, 2026).

### 4.2.2. Results

IO1 is solved using Algorithm 3 with binding tolerance $\epsilon_b = 1 \times 10^{-5}$, floor of $\varepsilon = 1 \times 10^{-6}$, premium bound $\kappa = 1$ (max observed link toll), maximum of 40 iterations for Algorithm 2', and up to maximum number of iterations for L-BFGS-B of $N_{clear} = 20{,}000$. The runs takes 189 seconds on average (144 – 266 s). Estimation itself is fast: the buyer QP of step 4 solves in 0.10 – 0.32 s, growing roughly linearly with the number of observed cells (805 – 2,599 active cells and 2,414 – 12,340 imposed zero cells from 10% to 50% sampling), and the seller QP is negligible. The remaining time is spent in the evaluation solves of steps 5 – 6. Algorithm 2' converges in 18 – 22 iterations in all 25 runs, and the clearing solve in 8,700 – 10,200 L-BFGS-B iterations. Run time does not increase with the sampling rate.

Two metrics are tracked across sampling rates: the correlation between modeled and actual $\boldsymbol{\sigma}$; and the weighted mean absolute percent error (wMAPE) against the held-out links. All are measured against ground truth values known to the oracle.

For each metric $x_n$, the mean $\bar{x}$ across the five runs are computed at each sampling rate, as well as the sample standard deviation $s$, which is used to determine a 90% confidence interval of the sampling design (these intervals reflect variability over the sampling design, not uncertainty about the true values) using the student t-distribution of $\bar{x} \pm 2.132 \frac{s}{\sqrt{5}}$. These results are presented in Fig. 7.

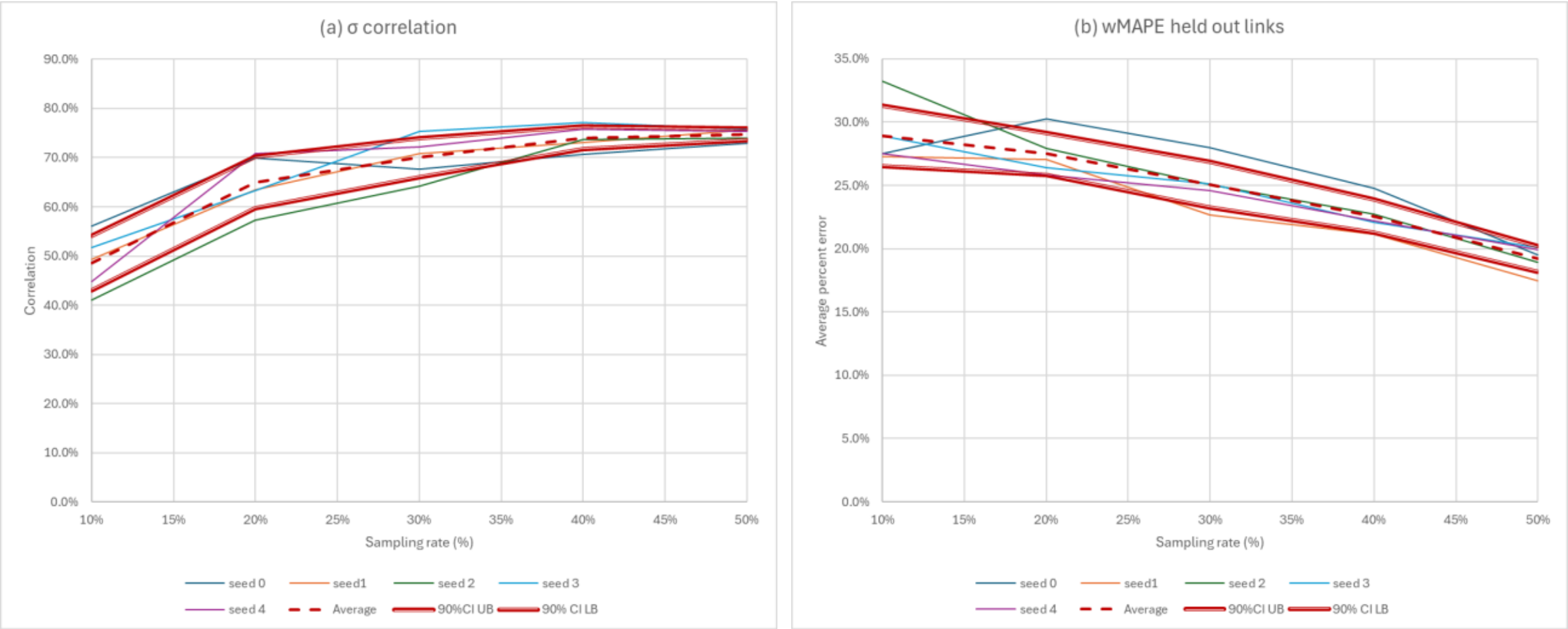

Fig. 7. Summary of results of 5 seeds of IO1 solved using Algorithm 3, under sampling rates 10% - 50% with average and 90% confidence intervals of the estimator over the sampling design: (a) correlation of modeled $\sigma_a$ compared to actual (including unobserved links); (b) weighted mean absolute percent error against held-out links.

The results are promising. In Fig. 7(a), we see the correlation between the modeled and actual rationing prices steadily growing from 50% at 10% sampling to plateau around 75% at 50% sampling. This suggests that while flow fit is present at 10% sampling, allocating the correct prices can continue improving with higher sampling rates.

In Fig. 7(b), the wMAPE shows the percent error decreasing from 29% at 10% sampling down to under 20% at 50% sampling. The present error is due to the misspecification of the model, but we can also see that as the sampling increases, the error of the held-out links generally decreases. A summary of other statistics at the 50% sampling are shown in Table 2.

The flow fit is high: the estimated model reproduces the oracle link flows with $R^2$ between 0.96 – 0.98 at every sampling rate, even though it omits the 97 station constants of the data-generating model. A good flow fit does not, however, imply a correct welfare level. $\hat{G}(\boldsymbol{\tau}^{\mathbf{0}})$ is underestimated by 17-33%, while $\hat{G}(\boldsymbol{\tau}^{*})$ is within 6% of the truth. The omitted station constants are absorbed mainly by the drive constant on the outside links, estimated at -0.5 to -1.6 against a true +5.54. This shifts the level of the travelers' surplus at the observed fares without changing which routes they use. Flows identify utility differences between alternatives, not the level of utility relative to the outside option. Flow-fit statistics alone are therefore not a sufficient validation for welfare analysis, and the clearing solution $\hat{G}(\boldsymbol{\tau}^*)$, which depends mainly on the relative attractiveness of the system's links, is the more robust welfare output.

Table 2. Summary of results of the 5 seeds at 50% sampling

| **Metric** | **Average across 5 seeds at 50% sampling (% error)** |
|---|---|
| $R^2$ | 96% |
| Gross surplus $\hat{G}(\tau^0)$ | \$596,007 (-33.4%) |
| Gross surplus at social optimum $\hat{G}(\tau^*)$ | \$1,851,898 (-5.8%) |
| Deadweight loss (DWL) | \$1,255,892 (+17.3%) |
| Deadweight loss (DWL) as % of $\hat{G}(\tau^*)$ | 67.8% vs 54.5% (+13.3 pp) |
| Absolute mean pricing wedge $\overline{\lvert\Delta_a\rvert}$ | \$1.63 (-33.6%) |
| Algorithm 3 estimation time (buyer QP, steps 1-4) | 0.31 s |
| Total run time, including P2 evaluation solves (steps 5-6) | 167 s |

To dig into the significance of these performances, we further conduct an ablation study of three factors related to Algorithm 3. The first is what if the only the inactive cells reachable from a cycle projection can be used for the constraints instead of all the observable inactive cells. This test evaluates the information loss from adopting a cycle projection approach similar to Fosgerau et al. (2026). The second factor considered is if all the inactive cells are removed entirely from consideration, which considers the value considered by Rosenblad et al. (2026). The third factor is ignoring the capacity rationing entirely by forcing $\sigma = 0$. The three factors are evaluated using two metrics: wMAPE of the held-out links, and the correlation between the pricing wedge for each link for sample rates 10%/30%/50% as shown in Table 3. The latter metric is evaluated instead of the $\sigma$ correlation since this would be unavailable for the "No $\sigma$" scenario.

Table 3. Ablation study of impacts of accounting for projection-limited inactive cells, all inactive cells, and no rationing prices

| **Sample rate** | **wMAPE held-out links** | | | | **Wedge correlation** | | | |
|---|---|---|---|---|---|---|---|---|
| | **Algorithm 3** | **Cycle projections** | **No inactive** | **No $\boldsymbol{\sigma}$** | **Algorithm 3** | **Cycle projections** | **No inactive** | **No $\boldsymbol{\sigma}$** |
| 10% | 28.9% | 34.7% | 64.7% | 33.3% | 80% | 62% | 69% | 73% |
| 30% | 25.1% | 32.3% | 57.7% | 29.3% | 75% | 53% | 44% | 71% |
| 50% | 19.2% | 20.9% | 44.6% | 22.5% | 76% | 61% | 40% | 72% |

The results suggest that the inactive-cell contribution is much more important to account for than the $\boldsymbol{\sigma}$ in the fixed price equilibrium. Even limiting inactive cells to only reachable links from cycle projection, the performance degrades to similar levels as having no $\boldsymbol{\sigma}$. When the inactive cells are completely removed, the wMAPE at the three sampling levels degrade by more than double from Algorithm 3. These results align with that of Rosenblad et al. (2026) which suggest higher sampling requirement when no inactive cell information is used.

For Algorithm 3, we see that the wedge correlation has already reached a relative plateau from 10% sampling on as the correlation does not improve further. Nonetheless, by removing either the inactive cell constraints or the $\boldsymbol{\sigma}$ consideration, the correlation drops in both.

Finally, we examine the outcome of the optimal toll schedule compared to the published toll schedule. While the tolls are link-based in the model, the network zone structure is fairly sequential. As a result, we can remap the average of the $\boldsymbol{\tau}^*$ across the five seeds at 50% sampling by weighting by trips traversing from each zone to zone 1. The comparison in Fig. 8 shows the published zonal fares and what the model's optimal fares based on the synthetic ground truth would look like, as well as the optimal tolls under the oracle. In both cases of the model and oracle, the optimal tolls are significantly lower than the published fares. This is because the optimal toll under this example ignores capital costs of operations and only considers flows inbound to NYC. Having reverse-peak and other trips included would likely increase the scarcity premium on binding links. Between the two optimal toll results, the fares are generally similar with the model tending to underpredict the optimal tolls, likely due to the misspecification.

## 5. Conclusion

Operationalizing assignment games to study multi-operator mobility markets as two-sided markets requires a method to estimate the parameters. Following the literature on double entropy-regularized aggregate assignment game formulations to enable model identification, we extend the framework to a path-based aggregate network assignment game. In such a model, travelers choose paths to travel while operators owning one or more links choose the amount of capacity to allocate to each link. The model is shown to have a unique primal solution, and a unique dual solution under the condition of Theorem 1(2), due to the entropy regularization, and preserves the structure of *Galichon and Salanié (2022)*. We further establish the decentralized behavior of travelers despite the socially optimal setting of the problem through first best pricing. The model produces outputs that readily support analysis of network investment allocations such as hubs and corridors, market entry/exit, value of network and market elements, and impacts of operator investments on other operators. An algorithm is proposed to solve the model. We extend the model to a PURC-based formulation in which operators deploy capacity on services, covering line-based and link-based operators in one model, with a solution algorithm and a fixed-price variant whose equilibrium exists (Theorem 3).

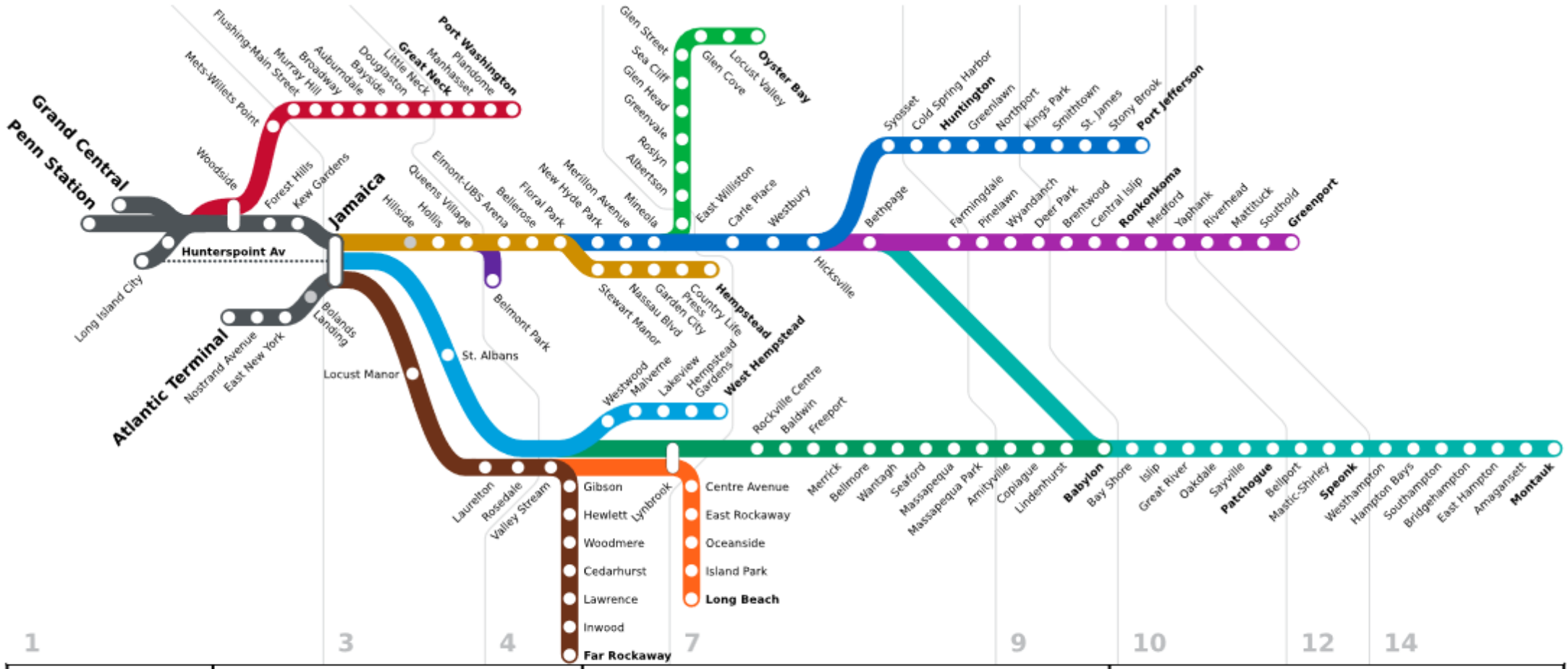

| From zone $i$ | Published fare to zone 1 ($\tau^0$) | Estimated optimal toll to zone 1 ($\hat{\tau}^*$) (% vs oracle) | Oracle optimal tolls to zone 1($\tau^*$) |
|---|---|---|---|
| 4 | $13.50 | $5.00 (-4.8%) | $5.25 |
| 7 | $15.25 | $6.25 (-24.2%) | $8.25 |
| 9 | $18.25 | $9.00 (-18.2%) | $11.00 |
| 10 | $21.50 | $13.50 (-16.9%) | $16.25 |
| 12 | $25.50 | $15.75 (-3.1%) | $16.25 |
| 14 | $33.00 | $19.75 (-2.5%) | $20.25 |

Fig. 8. Comparison of remapped tolls to zone fare schedule to compare published schedule against model optimal and oracle optimal schedules. LIRR Stations and Fare Zones map source: Wikipedia (2023)

An inverse optimization model is proposed for the PURC-based aggregate network assignment game. Conditions for identification are discussed. The inverse problem is a convex QP that separates into two blocks. The seller block recovers operating costs and group supply scales. The buyer block retains the node potentials, imposes every observed zero-flow cell, and supports rationing prices only on links observed at capacity.

Computational experiments conducted with a small toy network illustrate the forward models P1 and P2. A large-scale experiment with the LIRR network is used to evaluate the performance of the proposed inverse optimization algorithm, showing the algorithm can fit observed data well under an unknown specification ($R^2 \geq 0.96$), while the flow fit alone does not validate welfare levels. Ablation tests show the improvement due to the added design elements involving inactive cells and rationing pricing estimation. A comparison of the zonal fare schedules demonstrates how this model can be used to design fare schedules.

Future research will consider k-sided platforms, extend the pricing instruments to more complex subscription plans, and consider time dimensions with multiple periods. Opportunities for applications to different types of networks (airlines, freight, MaaS, humanitarian logistics, autonomous vehicle fleets, and vertically integrated energy-mobility systems) abound.

## Acknowledgments

Funding support from National Science Foundation CMMI-2423908 is gratefully acknowledged. Anthropic Claude was used to check derivations and to develop and debug the numerical implementation. The author verified all results and is responsible for the content.

## References

Agarwal, R., & Ergun, Ö. (2008). Mechanism design for a multicommodity flow game in service network alliances. *Operations Research Letters*, *36*(5), 520-524.

Agarwal, R., & Ergun, Ö. (2010). Network design and allocation mechanisms for carrier alliances in liner shipping. *Operations Research*, *58*(6), 1726-1742.

Ahuja, R. K., & Orlin, J. B. (2001). Inverse optimization. *Operations Research*, *49*(5), 771-783.

Becker, G. S. (1973). A theory of marriage: Part I. *Journal of Political Economy*, *81*(4), 813-846.

Becker, G. S. (1974). A theory of marriage: Part II. *Journal of Political Economy*, *82*(2, Part 2), S11-S26.

Bertsimas, D., Gupta, V., & Paschalidis, I. C. (2015). Data-driven estimation in equilibrium using inverse optimization. *Mathematical Programming*, *153*(2), 595-633.

Bie, J., & Lo, H. K. (2010). Stability and attraction domains of traffic equilibria in a day-to-day dynamical system formulation. *Transportation Research Part B: Methodological*, *44*(1), 90-107.

Boyce, D., Ralevic-Dekic, B., & Bar-Gera, H. (2004). Convergence of traffic assignments: how much is enough?. *Journal of Transportation Engineering*, *130*(1), 49-55.

Brander, J. A., & Zhang, A. (1990). Market conduct in the airline industry: an empirical investigation. *The RAND Journal of Economics*, 567-583.

Brucker, P., & Shakhlevich, N. V. (2009). Inverse scheduling with maximum lateness objective. *Journal of Scheduling*, *12*(5), 475-488.

Brueckner, J. K. (2001). The economics of international codesharing: an analysis of airline alliances. *international Journal of Industrial Organization*, *19*(10), 1475-1498.

Burkard, R. E., Pleschiutschnig, C., & Zhang, J. (2004). Inverse median problems. *Discrete Optimization*, *1*(1), 23-39.

Byrd, R. H., Lu, P., Nocedal, J., & Zhu, C. (1995). A limited memory algorithm for bound constrained optimization. *SIAM Journal on Scientific Computing*, *16*(5), 1190-1208.

Cantarella, G. E., & Watling, D. P. (2016). A general stochastic process for day-to-day dynamic traffic assignment: Formulation, asymptotic behaviour, and stability analysis. *Transportation Research Part B: Methodological*, 92, 3-21.

Carr, S. M., & Karmarkar, U. S. (2005). Competition in multiechelon assembly supply chains. *Management Science*, *51*(1), 45-59.

Cascetta, E., & Cantarella, G. E. (1991). A day-to-day and within-day dynamic stochastic assignment model. *Transportation Research Part A: General*, 25(5), 277-291.

Chen, L., Chen, Y., & Langevin, A. (2021). An inverse optimization approach for a capacitated vehicle routing problem. *European Journal of Operational Research*, *295*(3), 1087-1098.
Choo, E., & Siow, A. (2006). Who marries whom and why. *Journal of Political Economy*, *114*(1), 175-201.
Chow, J. Y. J. (2026). Aggregate Network Assignment Game Code and Data, https://github.com/BUILTNYU/AggregateNetworkAssignmentGame, last accessed 9/21/26.
Chow, J. Y. J., & Recker, W. W. (2012). Inverse optimization with endogenous arrival time constraints to calibrate the household activity pattern problem. *Transportation Research Part B: Methodological*, *46*(3), 463-479.
Chow, J. Y. J., Ritchie, S. G., & Jeong, K. (2014). Nonlinear inverse optimization for parameter estimation of commodity-vehicle-decoupled freight assignment. *Transportation Research Part E: Logistics and Transportation Review*, *67*, 71-91.
Dagsvik, J. K. (2000). Aggregation in matching markets. *International Economic Review*, *41*(1), 27-58.
Diamond, S., & Boyd, S. (2016). CVXPY: A Python-embedded modeling language for convex optimization. *Journal of machine learning research*, *17*(83), 1-5.
Djavadian, S., & Chow, J. Y. J. (2017). An agent-based day-to-day adjustment process for modeling 'Mobility as a Service' with a two-sided flexible transport market. *Transportation Research Part B: Methodological*, *104*, 36-57.
Facchinei, F., & Pang, J. S. (Eds.). (2003). *Finite-dimensional variational inequalities and complementarity problems*. New York, NY: Springer New York.
Fosgerau, M., Paulsen, M., & Rasmussen, T. K. (2022). A perturbed utility route choice model. *Transportation Research Part C: Emerging Technologies*, *136*, 103514.
Fosgerau, M., Nielsen, N., Rasmussen, T., & Yao, R. (2026). Estimating the perturbed utility route choice model with trip-level data. *arXiv preprint arXiv:2608.11464*.
Friesz, T. L., Gottfried, J. A., & Morlok, E. K. (1986). A sequential shipper-carrier network model for predicting freight flows. *Transportation Science*, *20*(2), 80-91.
FTA (2023). National Transit Summaries and Trends 2023. National Transit Database Website. https://www.transit.dot.gov/ntd/2023-national-transit-summaries-and-trends-ntst, last accessed 10/5/2026.
Gale, D., & Shapley, L. S. (1962). College admissions and the stability of marriage. *The American Mathematical Monthly*, *69*(1), 9-15.
Galichon, A., & Salanié, B. (2022). Cupid's invisible hand: Social surplus and identification in matching models. *The Review of Economic Studies*, *89*(5), 2600-2629.
Goulart, P. J., & Chen, Y. (2026). Clarabel: An interior-point solver for conic programs with quadratic objectives. *Mathematical Programming Computation*, 1-83.
Harker, P. T., & Friesz, T. L. (1986a). Prediction of intercity freight flows, I: theory. *Transportation Research Part B: Methodological*, *20*(2), 139-153.
Harker, P. T., & Friesz, T. L. (1986b). Prediction of intercity freight flows, II: mathematical formulations. *Transportation Research Part B: Methodological*, *20*(2), 155-174.
Hendricks, K., Piccione, M., & Tan, G. (1997). Entry and exit in hub-spoke networks. *The RAND Journal of Economics*, 291-303.
Kalai, E., & Zemel, E. (1982). Totally balanced games and games of flow. *Mathematics of Operations Research*, *7*(3), 476-478.
Krajewska, M. A., Kopfer, H., Laporte, G., Ropke, S., & Zaccour, G. (2008). Horizontal cooperation among freight carriers: request allocation and profit sharing. *Journal of the Operational Research Society*, *59*(11), 1483-1491.
Littlechild, S. C., & Owen, G. (1973). A simple expression for the Shapley value in a special case. *Management Science*, *20*(3), 370-372.
Liu, B., & Chow, J. Y. J. (2024). On-demand mobility-as-a-Service platform assignment games with guaranteed stable outcomes. *Transportation Research Part B: Methodological*, *188*, 103060.
Liu, B., Watling, D., & Chow, J. Y. J. (2025). Stochastic assignment games for Mobility-as-a-Service markets. *arXiv preprint arXiv:2512.19328*.
Liu, Z., & Yin, Y. (2026). Constructing transportation network equilibrium models from empirical data: An inverse learning perspective. *Transportation Research Part B: Methodological*, *214*, 103597.
MTA (2022). MTA Origin-Destination and Travel Surveys, https://data.ny.gov/Transportation/MTA-Origin-Destination-and-Travel-Surveys/hexe-btds/about_data, last accessed 9/12/2026.
MTA (2025). 2025 Annual Ridership Report, https://www.mta.info/document/205596, last accessed 9/12/2026.
MTA (2026). MTA/Long Island Rail Road Station Fares, Fare Chart: Ticket Types, effective August 20, 2023. https://www.mta.info/document/194866, last accessed 9/20/2026.

Nagurney, A., Dong, J., & Zhang, D. (2002). A supply chain network equilibrium model. *Transportation Research Part E: Logistics and Transportation Review*, *38*(5), 281-303.
Özener, O. Ö., & Ergun, Ö. (2008). Allocating costs in a collaborative transportation procurement network. *Transportation Science*, *42*(2), 146-165.
Pantelidis, T. P., Chow, J. Y. J., & Rasulkhani, S. (2020). A many-to-many assignment game and stable outcome algorithm to evaluate collaborative mobility-as-a-service platforms. *Transportation Research Part B: Methodological*, *140*, 79-100.
Park, J. H. (1997). The effects of airline alliances on markets and economic welfare. *Transportation Research Part E: Logistics and Transportation Review*, *33*(3), 181-195.
Rasulkhani, S., & Chow, J. Y. J. (2019). Route-cost-assignment with joint user and operator behavior as a many-to-one stable matching assignment game. *Transportation Research Part B: Methodological*, *124*, 60-81.
Reiss, P. C., & Spiller, P. T. (1989). Competition and entry in small airline markets. *The Journal of Law and Economics*, *32*(2, Part 2), S179-S202.
Rosenblad, B., Lin, X., & Yin, Y. (2026). Inverse Estimation of The Perturbed Utility Route Choice Model. *Available at SSRN 6756317*.
Shapley, L. S., & Shubik, M. (1971). The assignment game I: The core. *International Journal of Game Theory*, *1*(1), 111-130.
Stuart, A. M., & Wolfram, M. T. (2020). Inverse optimal transport. *SIAM Journal on Applied Mathematics*, *80*(1), 599-619.
Tarantola, A. (2005). *Inverse problem theory and methods for model parameter estimation*. Society for industrial and applied mathematics.
Tebaldi, C., & West, M. (1998). Bayesian inference on network traffic using link count data. *Journal of the American Statistical Association*, *93*(442), 557-573.
U.S. Census Bureau (2025). *ny_od_main_JT00_2022* [Data set]. LEHD Origin-Destination Employment Statistics, Version 8.4. https://lehd.ces.census.gov/data/lodes, accessed 9/12/2026.
Wikipedia (2023). LIRR schematic, https://commons.wikimedia.org/wiki/File:LIRR_schematic.svg, last accessed 10/8/2026.
Willumsen, L.G. (1978). Estimation of an O-D Matrix from Traffic Counts – A Review. Working Paper. Institute of Transport Studies, University of Leeds, Leeds, UK.
Xi, H., Aussel, D., Liu, W., Waller, S. T., & Rey, D. (2024). Single-leader multi-follower games for the regulation of two-sided mobility-as-a-service markets. *European Journal of Operational Research*, 317(3), 718-736.
Yang, H., Sasaki, T., Iida, Y., & Asakura, Y. (1992). Estimation of origin-destination matrices from link traffic counts on congested networks. *Transportation Research Part B: Methodological*, *26*(6), 417-434.
Yang, H., Wong, S. C., & Wong, K. I. (2002). Demand–supply equilibrium of taxi services in a network under competition and regulation. *Transportation Research Part B: Methodological*, *36*(9), 799-819.
Yang, H., & Chow, J. Y. J. (2026). Bilevel subsidy-enabled mobility hub network design with perturbed utility coalitional choice-based assignment. *Transportation Research Part A: Policy and Practice*, *208*, 104972.
Yao, R., Ma, X., & Zhang, K. (2026). Mobility-as-a-service (MaaS) system as a multi-leader-multi-follower game: A single-level variational inequality (VI) formulation. *Transportation Research Part C: Emerging Technologies*, *188*, 105703.
You, S. I., Chow, J. Y. J., & Ritchie, S. G. (2016). Inverse vehicle routing for activity-based urban freight forecast modeling and city logistics. *Transportmetrica A: Transport Science*, *12*(7), 650-673.
Zattoni Scroccaro, P., van Beek, P., Mohajerin Esfahani, P., & Atasoy, B. (2025). Inverse optimization for routing problems. *Transportation Science*, *59*(2), 301-321.
Zha, L., Yin, Y., & Yang, H. (2016). Economic analysis of ride-sourcing markets. *Transportation Research Part C: Emerging Technologies*, *71*, 249-266.
Zhou, J., Lam, W. H., & Heydecker, B. G. (2005). The generalized Nash equilibrium model for oligopolistic transit market with elastic demand. *Transportation Research Part B: Methodological*, *39*(6), 519-544.
Zhou, Y., Yang, H., Ke, J., Wang, H., & Li, X. (2022). Competition and third-party platform-integration in ride-sourcing markets. *Transportation Research Part B: Methodological*, *159*, 76-103.